\documentclass[fleqn,a4paper,12pt]{article}
\usepackage{amssymb}
\usepackage{makeidx}
\usepackage{amsmath}
\usepackage{graphicx}
\usepackage{lscape}
\usepackage{a4}
\usepackage{epsfig}
\usepackage{graphicx}

 \newcommand{\newc}{\newcommand}
\newc{\beq}{\begin{equation}} \newc{\eeq}{\end{equation}}
\newc{\bea}{\begin{array}} \newc{\eea}{\end{array}}
\newc{\ri}{{\mathrm i}}
\newc{\bW}{{\mathbf W}}
\newc{\bR}{{\mathbf R}}
\newc{\bN}{{\mathbf N}}
\newc{\Psibar}{\overline\Psi}
\newc{\w}{{\bf w}}
\newc{\E}{{\mathbf{E}}}
\newc{\bp}{{\bf p}}
\newc{\ta}{\tilde a}
\newc{\bV}{{\bf V}}
\newc{\bfV}{{\bf V}}
\newc{\bfG}{{\bf G}}
\newc{\bx}{{\bf x}}
\newc{\bu}{{\bf u}}
\newc{\bP}{{\bf P}}
\newc{\bJ}{{\bf J}}
\newc{\bK}{{\bf K}}
\newc{\pd}{{\partial}}
\newc{\ti}{{\times}}
\newc{\bA}{{\bf A}}
\newc{\bE}{{\bf E}}
\newc{\bfn}{{\bf\nabla}}
\newc{\ho}{\hookrightarrow}
\newc{\ra}{\rightarrow}
\newc{\bv}{{\bf v}}
\newc{\bb}{{\bf b}}
\newc{\bc}{{\bf c}}
\newc{\bd}{{\bf d}}
\newc{\tbb}{\tilde{\bf b}}
\newc{\tbc}{\tilde{\bf c}}
\newc{\tbd}{\tilde{\bf d}}
\newc{\bz}{{\bf 0}}
\newc{\bun}{{\bf 1}}
\newc{\bL}{{\bf L}}
\newc{\bS}{{\bf S}}
\newc{\bB}{{\bf B}}
\newc{\br}{{\bf r}}
\newc{\sig}{{\mathbf\sigma}}
\newc{\eg}{{\it e.g.\ }}
\newc{\bpi}{{\mathbf\pi}}
\newc{\ie}{{\it i.e.\ }}
\newc{\etal}{{\it et al}}
\newcommand{\p}{\partial}

\catcode`@=11 \long
\def\@caption#1[#2]#3{\par\addcontentsline{\csname
ext@#1\endcsname}{#1} {\protect\numberline{\csname
the#1\endcsname}{\ignorespaces #2}} \begingroup \small
\@parboxrestore \@makecaption{\csname fnum@#1\endcsname}
{\ignorespaces #3}\par \endgroup} \catcode`@=12

\begin{document}

\allowdisplaybreaks
 \begin{titlepage} \vskip 2cm

\begin{center} {\Large\bf Invariant solutions for equations
of axion electrodynamics} \footnote{E-mail: {\tt
gaponova@imath.kiev.ua},\ \ {\tt nikitin@imath.kiev.ua} } \vskip 3cm
{\bf {A.G. Nikitin $^a$ \ and\ Oksana Kuriksha $^b$ } \vskip 5pt {\sl
$^a$Institute of Mathematics, National Academy of Sciences of
Ukraine,\\ 3 Tereshchenkivs'ka Street,  01601 Kyiv-4, Ukraine\\
$^b$Petro Mohyla Black Sea State University,\\
 10, 68 Desantnukiv Street,
 54003 Mukolaiv,
 Ukraine\\}}
\end{center}
\vskip .5cm \rm

\begin{abstract} Using the three-dimensional subalgebras of
the Lie algebra of Poincar\'e group an extended class of exact
solutions for the  field equations of the axion electrodynamics  is
obtained. These solutions include arbitrary parameters and arbitrary
functions as well. The most general solutions include six arbitrary
functions. Among them there are bound and square integrable
solutions which propagate faster than light. However, their energy velocities
 are smaller than the velocity of light.
\end{abstract}

\end{titlepage}

\setcounter{footnote}{0} \setcounter{page}{1}
\setcounter{section}{0}

\section{Introduction}

Group analysis  of partial differential equations is a well developed  branch of mathematics
 including a number of interesting fundamental problems. But maybe its the
 main
 value are the powerful tools
 for construction of exact solutions of complicated nonlinear
 equations. Sometimes it is the group analysis which gives  the only hope to obtain at least some solutions for
 complicated physical (chemical, biological...) models.

In this paper we present some results obtained with application
of the group analysis
 to the important physical model called axion
electrodynamics. There was a lot of motivations for this
research, both physical and mathematical.

Axions are hypotetical particles which belong to main candidates to
form the dark matter, see, e.g. \cite{raflet}.  Additional arguments
to investigate axionic theories appeared in solid states physics,
since the axionic-type interaction terms appears in the theoretical
description of crystalline solids called topological insulators
\cite{Qi}. The existence of a pseudoscalar (axion) field can be
extracted from the experimental data concerning electric field
induced magnetization on $\text{Cr}_2\text{O}_3$ crystals or the
magnetic field-induced polarization \cite{heht}. In addition, the
axion hypothesis makes it possible to resolve the fundamental
problem of quantum chromodynamics connected with the CP symmetry
violation in interquark interactions \cite{pec}--\cite{wilczek}.
Thus it is interesting to make group analysis of axionic theories
which are requested in the three fundamental branches of physics,
i.e.,  the cosmology, condensed matter physics and quantum
chromodynamics.

Let us present three more motivations with are very inspiring for us.
Recently a new exactly solvable model for neutral Dirac fermions had
been found and other integrable models for such particles had been
indicated \cite{ninni}. But these models involve the external
electromagnetic field which does not solve Maxwell equations with
physically reasonable currents. However, these fields solve
equations of axion electrodynamics.

We had classified exactly solvable quantum mechanical systems including
shape invariant matrix potentials \cite{NKa1}, \cite{NKa2}.
Some of these systems also include solutions of field equations of
axion electrodynamics.

We have described the finite dimensional indecomposable vector
representations of the homogeneous Galilei group and construct
Lagrangians which admit the corresponding symmetries
\cite{NN1}--\cite{NN3} . The Lagrangian of axion electrodynamics
appears to be the relativistic counterpart of some of
 these Galilei invariant Lagrangians \cite{NK1}.

In addition, axion electrodynamics is a nice and  complicated mathematical model which  needs good group-theoretical grounds. In this preprint we present such grounds and also  find an extended class of exact solutions for the related field equations.

\section{Equations of axion electrodynamics}
Let us consider the following generalized Lagrangian:
\beq\label{1}L=\frac12 p_\mu p^\mu-\frac14 F_{\mu\nu}F^{\mu\nu}+
\frac{\kappa}{4}\theta F_{\mu\nu}\tilde F^{\mu\nu}-V(\theta).\eeq
Here $F_{\mu\nu}=\p_\mu A_\nu-\p_\nu A_\mu$, $A_\mu$ is the
vector-potential of electromagnetic field, $\tilde
F_{\mu\nu}=\frac12 \varepsilon_{\mu\nu\rho\sigma}F^{\rho\sigma}$,
$\theta$ is the axion field, $p_\mu=\partial_\mu\theta$,  $V(\theta)
$ is a function of $\theta $ and $\kappa$ is a dimensionless
constant which is supposed to be nonzero and can be rescaled to the
unity.

If $\theta\equiv0$ and $V=0$ then formula (\ref{1}) gives the Lagrangian of Maxwell
field. For $V(\theta)=\frac12m^2\theta^2$ equation (\ref{1}) reduces to the
standard Lagrangian of axion electrodynamics.

The  Euler-Lagrange equations corresponding to Lagrangian (\ref{1})
have the following form:
\begin{gather}\begin{split}
&\nabla \cdot \textbf{E}=\kappa\textbf{p}\cdot\textbf{B},\\
&\partial_0 \textbf{E}-\nabla \times
\textbf{B}=\kappa(p_0\textbf{B}+\textbf{p}\times\textbf{E}),\\
&\nabla \cdot \textbf{B}=0,\label{kuriksha:PDE}\\
&\partial_0 \textbf{B}+\nabla \times \textbf{E}=0,\end{split}\\ \Box
\theta =-\kappa \textbf{E} \cdot \textbf{B}+F.
\label{kuriksha:PDE5}\end{gather} Here $\textbf{B}$ and $\textbf{E}$
are vectors of the magnetic and electric fields whose components are
expressed via $F^{ \mu\nu}$ as $ E^a=F^{0a},\quad
B^a=-\frac12\varepsilon^{0abc}F_{bc},$ and $F=-\frac{\p
V}{\p\theta},\ \ \Box=\p_0^2-\p_1^2-\p_2^2-\p_3^2,\quad
 \nabla^a=\p_a=\frac{\p}{\p x_a}$,
$a=\overline{1,3}$.

We will search for solutions of system
 (\ref{kuriksha:PDE}), (\ref{kuriksha:PDE5}) with $\kappa=1$, which can be done up to equivalence scaling  transformations if $\kappa\neq0$.   In
 addition, we will consider also the following system
\begin{gather}\label{eq3}\begin{split}&
\nabla \cdot \textbf{E}=\kappa\textbf{p}\cdot\textbf{E},\\&
\partial_0 \textbf{E}-\nabla \times
\textbf{B}=\kappa(p_0\textbf{E}-\textbf{p}\times\textbf{B}),\\&
\nabla \cdot \textbf{B}=0,\\&
\partial_0 \textbf{B}+\nabla \times \textbf{E}=0,
\\ &\Box \theta =\kappa (\textbf{B}^2 - \textbf{E}^2)
+F\end{split}\end{gather} which includes  a scalar  field $\theta$,
while in equations (\ref{kuriksha:PDE}) $\theta$ is a pseudoscalar.

We shall present symmetries of  equations (\ref{kuriksha:PDE}),
(\ref{kuriksha:PDE5}) and (\ref{eq3})  with arbitrary function
$F(\theta)$ and also their exact solutions, which can be found using
these symmetries. We shall present also some results of group
classification for more general systems with $F$ being an arbitrary
function of $\theta$ and $p_\mu p^\mu$.

\section{Exact solutions: definitions and examples}
\subsection{Algorithm for finding group solutions}
Since the system (\ref{kuriksha:PDE}), (\ref{kuriksha:PDE5})  admits
rather extended symmetries, it is possible to find a number of its
exact solutions. The algorithm for construction of group solutions
of partial differential equations goes back to Sophus Lie. Being
applied to system (\ref{kuriksha:PDE}), (\ref{kuriksha:PDE5}) it
includes the following steps (compare, e.g., with \cite{olver}):
\begin{itemize}
\item To find a basis of the maximal Lie algebra $A_m$ corresponding to
continuous local symmetries  of the equation.
\item To find the optimal system of subalgebras $SA_\mu$ of algebra $A_m$. In the
case of PDE with four independent variables like system
(\ref{kuriksha:PDE}), (\ref{kuriksha:PDE5})  it is reasonable to
restrict ourselves to three-dimensional subalgebras. Their basis
elements have the unified form
$Q_i=\xi^\mu_i\p_\mu+\varphi^k_i\p_{u_k}, \ i=1,2,3$ where $u_k$ are
dependent variables (in our case we can chose $u_a=E_a,\
u_{3+a}=B_a, \ u_7=\theta,\ a=1,2,3$).
\item Any three-dimensional subalgebra $SA_\mu$ whose basis elements
satisfy the conditions \beq\label{la}\texttt{rank}
\{\xi^\mu_i\}=\texttt{rank} \{\xi^\mu_i,\varphi^k_i\}\eeq and
\beq\label{aala}\texttt{rank} \{\xi^\mu_i\}=3\eeq
 gives rise
to change of variables which reduce system (\ref{kuriksha:PDE}),
(\ref{kuriksha:PDE5}) to a system of ordinary differential equation
(ODE). The new variables include all invariants of three parameter
Lie groups corresponding to the optimal subalgebras $SA_\mu$.
\item Solving if possible the obtained ODE one can generate an
exact (particular) solution of the initial PDE.
\item Applying to this solution the general symmetry group transformation
it is possible to generate a family of exact solutions depending on
additional arbitrary (transformation) parameters.
\end{itemize}

The first step of the algorithm presupposed finding the maximal Lie
symmetry of the considered systems. These symmetries are presemted
in the following subsection.

\subsection{Group classification of system (\ref{kuriksha:PDE}),
(\ref{kuriksha:PDE5}) }

The system of equations (\ref{kuriksha:PDE5}) includes the arbitrary
element $F(\theta)$ thus its  symmetries might be different for
different $F$. The group classification of these equation consists
in complete description of their continuous symmetries together with
the specification of all functions $F$ corresponding to different
symmetries.

It has been proven in \cite{NK1} that the  maximal continuous
invariance group of system (\ref{kuriksha:PDE}),
(\ref{kuriksha:PDE5}) with {\it arbitrary} function $F(\theta)$ is
the group P(1,3). The corresponding Lie algebra p(1,3) is spanned on
the following basis elements:
\begin{gather}\begin{split}&
P_0=\p_0, \quad P_a=\p_{a},\\& J_{ab}=x_a \p_{b}-x_b \p_{a}+B^a
\p_{B^b}-B^b \p_{B^a}+E^a\p_{E^b}-E^b \p_{E^a},\\& J_{0a}=x_0
\p_{a}+x_a \p_0+\varepsilon_{abc} \left(E^b \p_{B^c}-B^b
\p_{E^c}\right)\end{split}  \label{kuriksha:yadro_operators}
\end{gather}
where $\varepsilon_{abc}$ is the unit antisymmetric tensor,
$a,b,c=1,2,3$.

For some particular functions $F$, namely, for $F=0, \ F=c$ and
$F=b\exp(a\theta)$ the symmetry of system (\ref{kuriksha:PDE}),
(\ref{kuriksha:PDE5}) is more extended. The corresponding Lie
algebra includes the following additional basis elements:
\begin{gather}P_4=\p_{\theta}, \quad D=x_0 \p_0+x_i\p_{i}-B^i\p_{B^i}-E^i\p_{E^i}
\ \text{ if }\  F(\theta)=0,\label{kuriksha:rozsh_operators2}\\
P_4=\p_{\theta}\  \text{ if } \ F(\theta)=c, \label{kuriksha:rozsh_operators1}\\
X=aD-2P_4\ \text{ if }\ F(\theta)=b \texttt{e}^{a
\theta}.\label{kuriksha:rozsh_operators3}
\end{gather}

Using the standard algorithm of group classification (see, e.g.,
\cite{olver})  we can find symmetries of a more general system
(\ref{kuriksha:PDE}), (\ref{kuriksha:PDE5}) with arbitrary element
$F$ being a function of both $\theta$ and its derivatives $p_\mu$.
Restricting ourselves to the case of Poincar\'e-invariant systems we
find that $F$ can be an arbitrary function of $\theta$ and $p_\mu
p^\mu$. Moreover, all cases when this symmetry can be extended are
presented by the following formulae:
\begin{gather}F=\kappa p_\mu p^\mu, \label{0}\\
F=f(p_\mu p^\mu),\label{c}\\
F= \texttt{e}^{a \theta}f\left(p_\mu p^\mu \texttt{e}^{-a
\theta}\right)\label{exp}
\end{gather}
where $f(.)$ is an arbitrary function on the argument given in the
brackets and $\kappa$ is an arbitrary constant. Symmetry algebras of
system (\ref{kuriksha:PDE}), (\ref{kuriksha:PDE5}) where $F$ is a
function given by formulae (\ref{0}), (\ref{c}) and (\ref{exp})
include all generators (\ref{kuriksha:yadro_operators}) and
operators presented in (\ref{kuriksha:rozsh_operators2}),
(\ref{kuriksha:rozsh_operators1}) and
(\ref{kuriksha:rozsh_operators3}) correspondingly.

  Finally,
the group classification of equations (\ref{eq3}) gives the same
results: this system is invariant w.r.t. Poincar\'e group for
arbitrary $F$. System (\ref{eq3}) admits more extended symmetry in
the cases enumerated in equations
(\ref{kuriksha:rozsh_operators2})--(\ref{exp}).

\subsection{Optimal subalgebras}

Thus, to generate  exact solutions of system (\ref{kuriksha:PDE}),
(\ref{kuriksha:PDE5}) we can exploit its invariance w.r.t. the
Poincar\'e group whose generators are presented in equation (\ref
{kuriksha:yadro_operators}). The subalgebras of algebra p(1,3)
defined up to the group of internal automorphism has been  found for
the first time in paper \cite{belorus}. We use a more advanced
classification of these subalgebras proposed in \cite{patera}. In
accordance with \cite{patera} there exist 30 non-equivalent
three-dimensional subalgebras $A_1,\ A_2,\ \cdots A_{30}$ of algebra
p(1,3) which we present in the following formulae by specifying
their basis elements :  \begin{equation}\begin{array}{ll} A_1:\ \langle P_0,P_1,P_2\rangle
;&
A_2: \ \langle P_1,P_2,P_3\rangle ; \\  A_3: \  \langle P_0-P_3,P_1,P_2\rangle ; &
A_4: \ \langle J_{03}, P_1, P_2\rangle ;\\  A_5: \ \langle J_{03},
P_0-P_3, P_1\rangle ;& A_6: \ \langle J_{03}+\alpha P_2, P_0,
P_3\rangle ;\\ A_7: \ \langle J_{03}+\alpha P_2, P_0-P_3, P_1\rangle
;& A_8: \ \langle J_{12}, P_0, P_3\rangle ;\\ A_9: \ \langle
J_{12}+\alpha
P_0, P_1, P_2\rangle ;& A_{10}: \ \langle J_{12}+\alpha P_3, P_1, P_2\rangle ;\\
A_{11}: \ \langle J_{12}- P_0+P_3, P_1, P_2\rangle ;&  A_{12}:
\ \langle G_1, P_0-P_3, P_2\rangle ;\\A_{13}: \ \langle G_1,
P_0-P_3, P_1+\alpha P_2\rangle ;& A_{14}: \ \langle G_1+P_2,
P_0-P_3, P_1\rangle ;\\ A_{15}: \ \langle G_1-P_0, P_0-P_3,
P_2\rangle ;& A_{16}: \ \langle G_1+P_0, P_1+\alpha P_2,
P_0-P_3\rangle ;\\A_{17}: \ \langle J_{03}+\alpha J_{12}, P_0,
P_3\rangle ;& A_{18}: \ \langle \alpha J_{03}+
J_{12}, P_1, P_2\rangle ;\\A_{19}: \ \langle J_{12}, J_{03},
P_0-P_3\rangle ;&  A_{20}: \ \langle G_1, G_2,
P_0-P_3\rangle ;\\A_{21}: \ \langle G_1+P_2, G_2+\alpha P_1+\beta P_2, P_0-P_3\rangle ;&
A_{22}: \ \langle G_1,
G_2+P_1+\beta P_2, P_0-P_3\rangle ;\\ A_{23}: \ \langle G_1, G_2+ P_2, P_0-P_3\rangle ;&
A_{24}: \ \langle G_1, J_{03}, P_2\rangle ;\\ A_{25}: \ \langle
J_{03}+\alpha P_1+\beta P_2, G_1, P_0-P_3\rangle ;& A_{26}: \
\langle J_{12}-P_0+ P_3, G_1, G_2\rangle ;\\ A_{27}: \ \langle
J_{03}+\alpha J_{12}, G_1, G_2\rangle ;& A_{28}:\ \langle
G_1,G_2,J_{12}\rangle ; \\ A_{29}:\ \langle
J_{01},J_{02},J_{12}\rangle;& A_{30}:\ \langle
J_{12},J_{23},J_{31}\rangle . \end{array}\label{suba}\end{equation}
 Here $P_\mu$ and
$J_{\mu\nu}$ are generators given by relations
(\ref{kuriksha:yadro_operators}), $G_1=J_{01}-J_{13},\
G_2=J_{02}-J_{23},\ \alpha$ and $\beta$ are arbitrary parameters.

Using subalgebras (\ref{suba}) we can deduce exact solutions for
system (\ref{kuriksha:PDE}), (\ref{kuriksha:PDE5}). Notice that to
make an effective reduction using the Lie algorithm, we can use only
such subalgebras whose basis elements satisfy conditions (\ref{la}).
This condition is satisfied by basis element of algebras
 $A_{1}-A_{27}$ but is not satisfied by $A_{28}, A_{29}, A_{30}$ and
 $A_6$ with $\alpha=0$. Nevertheless, the latter symmetries
 also can be used to generate  exact solutions in frames of the weak
 transversality
 approach discussed in \cite{wint}. We will use also a certain
 generalization of this approach.

 In the following sections we present the complete list of reductions and find exact
 solutions for system (\ref{kuriksha:PDE}), (\ref{kuriksha:PDE5}) which can be obtained
 using reduction w.r.t. the subgroups of Poincar\'e group. We will
 find also some solutions whose existence is caused by symmetry of this system
 with respect to the extended Poincar\'e group.

 \subsection{Plane wave solutions}
 Let us find solutions of system (\ref{kuriksha:PDE}), (\ref{kuriksha:PDE5})  which are invariant
 w.r.t. subalgebras  $A_1, A_2$ and $A_3$. Basis elements of all these subalgebras can be represented in the
 following unified form
 \beq\label{uni}A:\ \ \langle P_1,P_2,k P_0+\varepsilon P_3\rangle \eeq where
 $\varepsilon$ and $k$ are
 parameters. Indeed, setting in (\ref{uni}) $\varepsilon=-k$ we come to
 algebra $A_3$, for  $\varepsilon^2<k^2$ or  $k^2<\varepsilon^2$ algebra (\ref{uni})
 is equivalent to $A_1$ or $A_2$ correspondingly.

 Starting from this point we mark the components of vectors
 $\textbf{B}$  and $\textbf{E}$ by subindices, i.e.,  as
 $\textbf{B}=(B_1,B_2,B_3)$ and $\textbf{E}=(E_1,E_2,E_3)$.

 To find the related invariant solutions
 we need invariants of the groups generated by algebras
 (\ref{uni}). These invariants include the dependent variables
 $E_a, B_a, \theta$ ($a=1,2,3$) and independent variable
 $\omega=\varepsilon x_0
-kx_3$. Let us search for solutions of (\ref{kuriksha:PDE}),
(\ref{kuriksha:PDE5}) which are functions of $\omega$. Then
equations (\ref{kuriksha:PDE}) are reduced to the following system:
\beq\label{red}\bea{l}{\dot B}_3=0,\ {\dot E}_3={\dot \theta}B_3,\
 k{\dot E}_2=-\varepsilon{\dot B}_1,\ \ k{\dot E}_1=\varepsilon{\dot B}_2,\\
  \varepsilon{\dot
E}_1-k{\dot B}_2= {\dot \theta}(k E_2+\varepsilon B_1),\  k{\dot
B}_1+\varepsilon{\dot E}_2={\dot \theta}(\varepsilon{ B}_2-
kE_1)\eea\eeq where ${\dot B}_3=\frac{\p B_3}{\p \omega}$.

The system (\ref{red}) is easily integrated. If
$\varepsilon^2=k^2\neq0$ then
\begin{gather}\label{bub}E_1=\frac{\varepsilon}
k B_2=F_1,\ E_2=-\frac\varepsilon k B_1=F_2, \ E_3=c\theta +b,\
B_3=c\end{gather} where $F_1$ and $F_2$ are arbitrary functions of
$\omega$ while  $c$ and $b$ are arbitrary real numbers. The
corresponding equation (\ref{kuriksha:PDE5}) is reduced to the form
$e^2\theta=F(\theta)-be$, i.e., $\theta$ is proportional to
$F(\theta)-be$ if $e\neq0$. If both $e$ and $F$ equal to zero then
$\theta$ is an arbitrary function of $\omega$.

For $\varepsilon^2\neq k^2$ solutions of (\ref{red}) have the
following form:
 \begin{gather}\begin{split}\label{bed}&B_1=kc_1\theta-kb_1+\varepsilon c_2,\
 \  B_2=kc_2\theta-kb_2-\varepsilon c_1,\ \
B_3=c_3,\\& E_1=\varepsilon c_2\theta-\varepsilon b_2 -kc_1, \ \
E_2=-\varepsilon c_1\theta +\varepsilon b_1-kc_2,\ \
E_3=c_3\theta-b_3(\varepsilon^2-k^2)\end{split}\end{gather} where
$b_a$ and $c_a$ ($a=1,2,3$) are arbitrary constants. The
corresponding equation (\ref{kuriksha:PDE5}) takes the form
\beq\label{N3}{\ddot\theta}=-b\theta +c+\tilde F\eeq where
$b=\left(c_1^2+c_2^2+\frac{c_3^2}{\varepsilon^2- k^2}\right), \tilde
F=\frac{F}{(\varepsilon^2-k^2)}$ and $c=c_1b_1+c_2b_2+c_3b_3$.

  If $F=0$ or $F=-m^2\theta$
then (\ref{N3}) is reduced to the linear equation:
\beq\label{A00}{\ddot\theta}=-a\theta +c\eeq where
$a=c_1^2+c_2^2+\frac{c_3^2+m^2}{\varepsilon^2-k^2}.$ Thus
\begin{gather}\label{A1a}\theta=a_\mu \cos \mu\omega+b_\mu
\sin\mu\omega +\frac{c}{\mu^2}, \ \texttt{ if }
a=\mu^2>0,\\\label{A1b} \varphi=a_\sigma \texttt{e}^{\sigma
\omega}+b_\sigma \texttt{e}^{- \sigma\omega}-\frac{c}{\sigma^2}\
\texttt{ if } a=-\nu^2<0,\\\label{A1c} \theta=\frac12c
\omega^2+c_1\omega +c_2\ \ \ \text{ if }\ \ a=0\end{gather} where
$a_\mu, b_\mu, a_\sigma,  b_\sigma, c_1$ and $c_2$ are arbitrary
constants.

 One more plane wave solution of equations (\ref{kuriksha:PDE}),
(\ref{kuriksha:PDE5}) with $\kappa=1$ and $F=0$ can be written as:
\begin{gather}\label{wond1}\begin{split}&E_1=c_k\varepsilon\sin\omega-
d_k\varepsilon\cos\omega, \quad
E_2=c_k\varepsilon\cos\omega+d_k\varepsilon\sin\omega,
\\&B_1=-c_kk\cos\omega-d_kk\sin\omega,\quad
B_2=c_kk\sin\omega-d_kk\cos\omega,
 \\ &E_3=e,\ B_3=0,\ \ \theta=\alpha x_0-\nu
x_3+c_3,\quad \omega=\varepsilon x_0-kx_3\end{split}\end{gather}
where $e, c_k, d_k, \varepsilon, k, \alpha, \nu$ are arbitrary
constants restricted by the only constraint:
\begin{gather}\label{wond2}\varepsilon^2-k^2=\nu\varepsilon-
\alpha k.\end{gather}

If $\varepsilon=k$ then $\alpha=\nu$ and formulae (\ref{wond1})
present solutions depending on one light cone variable $x_0-x_3$.
However, for $\varepsilon\neq k$ we have solutions depending on two
different plane wave variables, i.e., $\varepsilon x_0-kx_3$ and
$\alpha x_0-\nu x_3$.

It is interesting to note that for fixed parameters $\alpha$ and
$\nu$ solutions (\ref{wond1}) for $E_a$ and $B_a$ satisfy the
superposition principle, i.e., a sum of solutions  with different
$\varepsilon, k, c_k$ and $d_k$ is also a solution of equations
(\ref{kuriksha:PDE}), (\ref{kuriksha:PDE5}) with $\kappa=1$ and
$F=0$.

Using symmetries of system (\ref{kuriksha:PDE}),
(\ref{kuriksha:PDE5}) it is possible to extend the obtained
solutions. Indeed, applying to (\ref{bub}), (\ref{bed}) the rotation
transformations \beq\label{N31}E_a\to E'_a=R_{ab}E_b, \ B_a\to
B'_a=R_{ab}B_b,\eeq where $\{R_{ab}\}$ is an arbitrary orthogonal
matrix of dimension $3\times3$, and then the Lorentz transformations
\begin{gather}\bea{l}E'_a\to
E'_a\cosh\lambda-\varepsilon_{abc}\lambda_bB'_c\frac{\sinh\lambda}{\lambda}
+\lambda_a\lambda_bE'_b\frac{1-\cosh\lambda}{\lambda^2},\\\\ B'_a\to
B'_a\cosh\lambda+\varepsilon_{abc}\lambda_bE'_c\frac{\sinh\lambda}{\lambda}
+\lambda_a\lambda_bB'_b\frac{1-\cosh\lambda}{\lambda^2},\ \ \
\lambda=\sqrt{\lambda_1^2+\lambda_2^2+\lambda_3^2}\eea
\label{N32}\end{gather} and transforming $\omega\to n_\mu x^\mu$
where $n_\mu$ are components of the constant  vector given by the
following relations:
\begin{gather}\label{rolo}\begin{split}&n_0=\cosh\lambda-\nu\lambda_bR_{b3}\frac{\sinh\lambda}
{\lambda},\\& n_a=\nu
R_{a3}-\lambda_a\frac{\sinh\lambda}{\lambda}-\nu\lambda_a\lambda_bR_{b3}
\frac{(1-\cosh\lambda)}{\lambda^2},\end{split} \end{gather}   we
obtain more general solutions of equations (\ref{kuriksha:PDE}),
(\ref{kuriksha:PDE5}).

    In formulae (\ref{N31})--(\ref{rolo}) summation is imposed over
    the repeated index $b, \ b=1,2,3$. Transformations (\ref{N31}) and
    (\ref{N32}) can be used also for other solutions presented in
    the  following text.

\subsection{Selected radial and cylindric solutions}

Let us present some other solutions of equations
(\ref{kuriksha:PDE}), (\ref{kuriksha:PDE5})  which can be
interesting from the physical point of view.

First we consider solutions which include the field of point charge,
i.e. \beq\label{N14} \displaystyle E_a=q\frac{x_a}{r^3},\
a=1,2,3\eeq where $r=\sqrt{x_1^2+x_2^2+x_3^2}$ and $q$ is a coupling
constant. Notice that up to scaling the dependent variables $x_a$ we
can restrict ourselves to $q=1$. The related vector $B_a$ is
trivial, i.e., $B_a=0$, while for  $\theta$ there are two solutions:
\beq\label{N15}
  \theta=
\frac{c_ax_a}{r^3}\ \texttt { and }\theta=
\frac{1}{r}\left(\varphi_1(x_0+r)+\varphi_2(x_0-r)\right) \eeq where
$\varphi_1$ and $\varphi_2$ are arbitrary functions of $x_0+r$ and
$x_0-r$ correspondingly, $c_a$ are arbitrary constants and summation
is imposed over the repeating indices $a=1,2,3$. These solutions
correspond to trivial nonlinear terms in (\ref{kuriksha:PDE}),
(\ref{kuriksha:PDE5}).

Radial solutions which generate nontrivial terms in the r.h.s. of
equations (\ref{kuriksha:PDE}), (\ref{kuriksha:PDE5}) with
$F=-m^2\theta$ can be found in the following form:
\begin{gather}\label{ha}{B}_a=-\frac{q{x}_a}{r^3}, \quad {E}_a=-\frac{q \theta {x}_a}{r^3},\ \
\theta=c_1\sin(mx_0)\texttt{e}^{-\frac{q}{r}}
\end{gather} where $c_1$ and $q>0$ are
arbitrary parameters. The components of magnetic field $B_a$ are
singular at $r=0$ while $E_a$ and $\theta$ are bounded for $0\leq
r\leq\infty$.

Solutions (\ref{N14})--(\ref{ha}) where obtained with using
invariants of algebra $A_{30}$.

 Let us present solutions which depend on two spatial variables but
 are rather
similar to the three dimensional Coulomb field. We denote
$x=\sqrt{x_1^2+x_2^2}$, then functions
\begin{gather} \label{N11}  E_1=-B_2=\frac{x_1}{x^3},\ E_3=0,\  B_1=E_2=\frac{x_2}{x^3}, \
 B_3=b,\
\theta=\arctan\left(\frac{x_2}{x_1}\right)\end{gather} where $b$ is
a  number, solve equations (\ref{kuriksha:PDE}),
(\ref{kuriksha:PDE5}) with $\kappa=1$ and $F=0$.

A particularity of solutions (\ref{N11}) is that, in spite of their
cylindric nature, the related electric field decreases with growing
of $x$ as the field of point charge in the three dimensional space.

Functions (\ref{N11}) solve the standard Maxwell equations with
charges and currents also. However, they correspond to the charge
and current densities proportional to $ 1/{x^3}$ which looks rather
nonphysical. In contrary, these vectors present consistent solutions
for equations of axion electrodynamics with zero axion mass.

Solutions (\ref{N11}) can be expressed via invariants of  the subgroup of the {\it
extended} Poincar\'e group whose Lie algebra is spanned on the basis
$\langle P_0,\ P_3,\ J_{12}+ P_4\rangle $, see equations
(\ref{kuriksha:yadro_operators}), (\ref{kuriksha:rozsh_operators1})
for definitions.

Let us write one more solution of equations (\ref{kuriksha:PDE}),
(\ref{kuriksha:PDE5}) with $F=0$:
\begin{gather} B_1=\frac{x_1x_3}{r^2x},\
B_2=\frac{x_2x_3}{r^2x},\ B_3=-\frac{x}{r^2},\ \theta=\arctan\left(\frac{x}{x_3}\right),\label{N20} \\
E_a=\frac{x_a}{r^2},\ a=1,2,3 \label{N21}\end{gather} where
$r=\sqrt{x_1^2+x_2^2+x_3^2},\ x=\sqrt{x_1^2+x_2^2}.$ The electric
field (\ref{N21}) is directed like the three dimensional field of
point charge but its strength is proportional to $1/r$ instead of
$1/r^2$.

Notice that functions (\ref{N20}), (\ref{N21}) solve equations
(\ref{eq3}) with $\kappa=1, F=0$ also.  Two additional stationary exact
solutions for these equations can be written
as:\beq\label{N12}\displaystyle E_a= \frac{x_a}{r^2},\ a=1,2,3; \
B_a=0, \ \theta=\ln (r)\eeq and\beq\label{N115}\displaystyle E_a=
\frac{x_a}{r},\ B_a=b_a,\ \ \theta=\ln (r)\eeq where $b_a$ are
constants satisfying the condition $b_1^2+b_2^2+b_3^2=1$. Functions
(\ref{N115}) solve equations (\ref{eq3}) with $F=0$ for $0<r<\infty$
while formula (\ref{N12}) gives solutions of equation (\ref{eq3})
with $F=p_ap^a$.

The complete list of exact solutions for equations
(\ref{kuriksha:PDE}), (\ref{kuriksha:PDE5})  obtained using
symmetries w.r.t. the 3-dimensional subalgebras of the Poincar\'e
algebra is presented in the following section.

\section{Complete list of invariant solutions}

In this section we present all exact solutions for equations
(\ref{kuriksha:PDE}), (\ref{kuriksha:PDE5}) which can be obtained
using symmetries w.r.t. the 3-dimensional subalgebras of the
Poincar\'e algebra. Basis elements of these subalgebras are given by
relations (\ref{suba}).

 We shall consider equations (\ref{kuriksha:PDE}), (\ref{kuriksha:PDE5}) with the
most popular form of function $F$, i.e., $F=-m^2\theta$, which is
the standard choice in axion electrodynamics. In addition, up to
scaling the dependent variables, we can restrict ourselves to the
case $\kappa=1$. Under these conventions the system
(\ref{kuriksha:PDE}), (\ref{kuriksha:PDE5}) can be rewritten in the
following form:
\begin{gather}\begin{split}&
\nabla \cdot \textbf{E}=\textbf{p}\cdot\textbf{B},\\&
\partial_0 \textbf{E}-\nabla \times
\textbf{B}=p_0\textbf{B}+\textbf{p}\times\textbf{E},\\& \nabla \cdot
\textbf{B}=0,\quad
\partial_0 \textbf{B}+\nabla \times \textbf{E}=0,\end{split}\label
{kuriksha:PDE3} \\
\label{kuriksha:PDE4}\Box \theta = -\textbf{E} \cdot
\textbf{B}-m^2\theta.\end{gather}

 In the following we present exact
solutions just for equations (\ref{kuriksha:PDE3}),
(\ref{kuriksha:PDE4}) for both nonzero and zero $m$.

 Solutions corresponding to algebras $A_1-A_3$ have been
 discussed in the previous subsection. Here we apply the remaining
 subalgebras from the list (\ref{suba}), grouping them into classes
 which correspond to similar reduced equations.

 \subsection{Reductions to algebraic equations}
Let us consider subalgebras $A_{11}, A_{20}, A_{26}$ and show that
using their invariants the system (\ref{kuriksha:PDE3}),
(\ref{kuriksha:PDE4}) can be reduced to algebraic equations.

Algebra $A_{11}: \quad \langle J_{12}- P_0+P_3, P_1, P_2 \rangle$.

Invariants $I$ of the corresponding Lie group are functions of the
dependent and independent variables involved into system
(\ref{kuriksha:PDE3}), (\ref{kuriksha:PDE4}), which satisfy the
following conditions
\begin{gather}\label{sp}P_1I=P_2I=0,\ \ (J_{12}-
P_0+P_3)I=0.\end{gather} The system (\ref{sp}) is non-degenerated
thus there are eight invariants which we choose in the following
form:
\begin{gather}\label{sp1}\begin{split}I_1=E_1\sin\zeta-E_2\cos\zeta,\
I_2=E_2\sin\zeta+E_1\cos\zeta,\ \\
I_3=B_1\sin\zeta-B_2\cos\zeta,\
I_4=B_2\sin\zeta+B_1\cos\zeta,\\
I_5=E_3,\ I_6=B_3, \ I_7=\theta, \ I_8=\omega=x_0+x_3\ \ \ \ \ \
\end{split}\end{gather} where $\zeta=\frac12({x_3-x_0})$ and
$I_\alpha,\ \alpha=1,2,...7$ are arbitrary functions of $\omega$.
Solving (\ref{sp1}) for $E_a, B_a$ and $\theta$ and using
(\ref{kuriksha:PDE3}) we obtain
\begin{gather}\label{ar4}
E_1=B_2=c_1 \sin {\zeta}+c_2 \cos {\zeta},\quad E_2=-B_1=c_2 \sin
{\zeta}-c_1 \cos {\zeta},\\ E_3=c_3\theta+c_4,\ \ B_3=c_3, \
\end{gather}
where  $c_1,\ c_2,\ c_3,\ c_4$ are arbitrary real constants and
$\theta$ is a function of $\omega$ which, in accordance with
(\ref{kuriksha:PDE4}), should satisfy the following linear algebraic
relation: \beq\label{ar5} (c_3^2+m^2)\theta+c_3c_4=0.\eeq
 Thus $\theta=-\frac{c_3c_4}{c_3^2+m^2}$ if the sum in brackets  is nonzero and
 $\theta$ is an
arbitrary constant  provided $c_3=m=0$.

Notice that solution (\ref{ar4}) can be generalized to the following one:
\begin{gather}\begin{split}&E_1=B_2=f(\zeta), \ E_2=-B_1=g(\zeta), \  E_3=c_3\theta+c_4,\ H_3=c_3
\end{split}\label{ar6}\end{gather}
where $f(\zeta)$, $g(\zeta)$ are arbitrary functions and $\theta$
again is defined by equation (\ref{ar5}). However, solution
(\ref{ar6}) cannot be obtained via symmetry reduction.

In analogous way we obtain solutions corresponding to subalgebras $
A_{20} \texttt{ and } A_{26}$.
 Algebra  $A_{20}: \quad \langle
G_1, G_2, P_0-P_3 \rangle$,
\begin{gather*}
B_1=E_2-\frac{c_2}{\omega}=\frac{-2 c_1 x_1
x_2+c_2(x_1^2-x_2^2)+2c_3x_1+2c_3x_2\theta+2c_4x_2}
{2\omega^3}+\varphi_1,\\
B_2=-E_1+\frac{c_1}{\omega}=\frac{c_1(x_1^2-x_2^2)+2c_2x_1x_2+2c_3x_2-2c_3x_1\theta-2c_4x_1}{2\omega^3}+\varphi_2,\\
B_3=\frac{-c_1x_2+c_2x_1+c_3}{\omega^2},\ \
E_3=\frac{-c_1x_1-c_2x_2+c_3 \theta+c_4}{\omega^2}
\end{gather*}
where $\varphi_i$ are functions of $\omega=x_0+x_3, $
\begin{gather*}
\theta=-\frac{(c_1\varphi_1
+c_2\varphi_2)\omega^3+c_3c_4}{c_3^2+m^2\omega^4}\ \ \   \text{ if
}\ \ c_3^2+m^2>0;\\\theta=\varphi_3,\ c_1\varphi_1+c_2\varphi_2=0\ \
\ \text{ if  }\ \ c_3^2+m^2=0.
\end{gather*}
Algebra $ A_{26}: \quad \langle J_{12}-P_0+ P_3, G_1, G_2 \rangle$,
\begin{gather*}
B_1=\frac{c_1x_1x_2}{\omega^3}\cos{\zeta}+\frac{c_2
x_2}{\omega^3}+\frac{c_1\left((\dot\theta-2)\omega^2+
2(x_1^2-x_2^2)\right)}{4\omega^3}
\sin {\zeta},\\
B_2=\frac{c_1x_1x_2}{\omega^3}\sin
{\zeta}+\frac{c_1\left((\dot\theta-2)\omega^2-
2(x_1^2-x_2^2)\right)}{4\omega^3}\cos{\zeta},\\
B_3=\frac{c_1x_1}{\omega^2}\sin {\zeta}-
\frac{c_2x_1}{\omega^3}+\frac{c_1x_2}{\omega^2}\cos{\zeta},\ \
E_1=-B_2-\frac{c_1}{\omega}\cos\zeta,\\
E_2=B_1+\frac{c_1}{\omega}\sin\zeta,\ \
E_3=\frac{c_1x_2}{\omega^2}\sin
{\zeta}+\frac{c_2}{\omega^2}+\frac{c_1x_1}{\omega^2}\cos{\zeta},\\
\theta=0, \ \ \texttt{if} \ \ m\neq0,\ \ \theta=\varphi(\omega)\ \
\texttt{if} \ \ m=0\end{gather*} where
$\zeta=\frac{x^2}{\omega}+\frac{\theta}{2},\
x^2=x_0^2-x_1^2-x_2^2-x_3^2$ and $\varphi(\omega)$ is an arbitrary
function of $\omega=x_0+x_3$.
\subsection{Reductions to linear ODE}

 The next class includes subalgebras $A_5, A_7, A_{15}, A_{16}$ and $A_{25}$.
  Using them we shall reduce the system (\ref{kuriksha:PDE3}),
  (\ref{kuriksha:PDE4}) to the only
linear ordinary differential equation (\ref{A00}).

Let us start with algebra $A_5$ whose basis elements are  $\langle
J_{03}, P_0-P_3, P_1 \rangle$. The corresponding invariant solutions
of equations (\ref{kuriksha:PDE3}), (\ref{kuriksha:PDE4}) have the
following form:
\begin{gather*}
B_1=E_2=(x_0+x_3)\left(c_1\theta+c_2\right),\quad
B_2=-E_1=c_1(x_0+x_3),\\
B_3=-c_3 \theta+c_4,\quad
 E_3=c_3,\ c_1c_2=0.\end{gather*}

 Function $\theta=\varphi(\omega)$ depends on the only variable $\omega=x_2$ and
 satisfies equation (\ref{A00}) where $a=c_3^2-m^2,\ c=c_3c_4$. Thus its possible
 forms
  are given by equations (\ref{A1a})--(\ref{A1c}).

 Algebra $A_7: \quad \langle
J_{03}+\alpha P_2, P_0-P_3, P_1 \rangle$
\begin{gather*}
B_1=E_2=\frac{ -c_1 \theta+c_2}{x_0+x_3}, \quad
B_2=-E_1=\frac{-\alpha c_3 \theta + \alpha c_4-c_1}{x_0+x_3}, \\
B_3=-c_3 \theta +c_4,\quad E_3=c_3.
\end{gather*}

Possible functions $\theta=\varphi(\omega)$ again are given by
equations (\ref{A1a})--(\ref{A1c}) where $a=c_3^2-m^2, c=c_3c_4$ and
$\omega=x_2-\alpha\ln{|x_0+x_3|}$.

Algebra $ A_{15}: \quad \langle G_1-P_0, P_0-P_3, P_2 \rangle$,
\begin{gather*}
B_1=E_2=-c_2 (x_0+x_3) \theta-c_1(x_0+x_3),\ \  B_3=c_2 \theta+c_1,\\
B_2=-E_1=c_3(2\omega-x_1)+c_2(x_0+x_3), \ \
 E_3=c_3(x_0+x_3)+c_2
\end{gather*}
where $\omega=x_1+\frac12(x_0+x_3)^2$.  Expressions for $\theta$ are
given by equations (\ref{A1b}), (\ref{A1c}) where
$\sigma^2=m^2+c_2^2, \ c=c_1c_2$.

Algebra $ A_{16}: \quad \langle G_1+P_0, P_1+\alpha P_2, P_0-P_3
\rangle$,
\begin{gather*}
B_1=(x_0+x_3)(c_3\theta-c_4)+
\frac12c_1(x_0+x_3)^2+\frac{c_5}{1+\alpha^2}(\theta-\alpha)+c_2,\\
B_2=c_1(\omega-\frac\alpha2(x_0+x_3)^2)+
c_3(x_0+x_3)+\frac{c_5}{1+\alpha^2}(\alpha\theta+1)+\alpha
c_2,\\
B_3=-c_3 \theta -c_1(x_0+x_3)+c_4,\
E_1=-B_2-{\alpha c_1},\\
E_2=B_1+{c_1},\ \ \ \ \ \  E_3=-\alpha c_1(x_0+x_3)+c_3
\end{gather*}
where $\omega=x_2-\alpha x_1-\frac{\alpha}2(x_0+x_3)^2,$
\begin{gather*} \theta=\frac1{\alpha^2+1}\left(\frac
{c_1^2}{6}\omega^3+\frac12(c_3c_4+c_1c_5)\omega^2\right)+c_7\omega+c_8
\ \texttt{ if } \ c_3^2= m^2,\\
\theta=\varphi+\frac{c_1^2\omega}{c_3^2-m^2} \ \texttt{ if } \
c_3^2\neq m^2.
 \end{gather*}
 Here $\varphi$ is the function of $\omega$ given by equations
 (\ref{A1a})--(\ref{A1c}) where $\mu^2=-\sigma^2=\frac{c_3^2-m^2}{\alpha^2+1},\
 c=\frac{c_3c_4+c_1c_5}{\alpha^2+1}$.

 Algebra $ A_{25}: \quad \langle J_{03}+\alpha
P_1+\beta P_2, G_1, P_0-P_3 \rangle$,
\begin{gather*}
B_1=E_2=\frac{c_3 +(c_3\theta+c_2)\zeta}{x_3+x_0},\quad
B_2=-E_1=\frac{\beta c_3\theta+c_1 +c_3 \zeta}{x_3+x_0},\\
B_3=c_3\theta+c_2,\ \
 E_3=-c_3,
\end{gather*}
where $\zeta=x_1-\alpha \ln{|x_3+x_0|}$ and $\theta=\varphi(\omega)$
is a function of $\omega=x_2-\beta \ln{|x_3+x_0|}$ given by
equations (\ref{A1a})--(\ref{A1c}) with $c=-c_2c_3$ and
$\mu^2=-\sigma^2=c_3^2-m^2$.

Consider now reductions which can be made with using invariants of
subalgebras $A_4, A_8, A_{19}, A_{24}$ and $ A_{27}$. In this way we
will reduce the system (\ref{kuriksha:PDE3}), (\ref{kuriksha:PDE4})
to linear ODEs which, however, differ from (\ref{A00}).

   Algebra $A_4:\ \ \ \langle J_{03}, P_1, P_2
\rangle$,
\begin{gather}\begin{split}
&B_1=\frac{-c_2x_3 \theta+c_6x_3-c_1x_0}{\omega^2},\ \
B_2=\frac{-c_1x_3 \theta+c_5x_3+c_2x_0}{\omega^2},\ \ B_3=c_3,
\\
&E_1=\frac{-c_1x_0 \theta+c_5x_0+c_2x_3}{\omega^2},\ \
E_2=\frac{c_2x_0 \theta+c_1x_3-c_6x_0}{\omega^2},\ \ E_3=c_3 \theta+c_4
\end{split}\label{soll1}
\end{gather}
where $c_1,\cdots,c_6$ are arbitrary constants,
$\theta=\theta(\omega)$ and $\omega^2=x_0^2-x_3^2$. Substituting
(\ref{soll1}) into (\ref{kuriksha:PDE4}) we obtain:
\beq\label{sol2}\omega^2\ddot\theta+\omega\dot\theta+
(\nu^2+\mu^2\omega^2)\theta=\delta+\alpha \omega^2\eeq where
$\nu^2=c_1^2+c_2^2, \ \mu^2=c_3^2+m^2,\  \delta=c_1c_5+c_2c_6,
\alpha=c_3c_4$ and $\dot\theta=\p\theta/\p\omega$.

The general real solution of equation (\ref{sol2}) for $x_0^2>x_3^2$
is:
\begin{gather}\theta=c_7\left(\texttt{J}_{\mathbf{i}\nu}(\mu \omega)+
\texttt{J}_{-\mathbf{i}\nu}(\mu
\omega)\right)+c_8\left(\texttt{Y}_{\mathbf{i}\nu}(\mu \omega)+
\texttt{Y}_{-\mathbf{i}\nu}(\mu
\omega)\right)\nonumber\\\label{WEB}+
\frac{\delta\pi}{2\nu}\left(\coth
\left(\frac{\pi\nu}{2}\right)J_{\texttt{i}\nu}
 (\mu \omega)+\texttt{i}E_{\texttt{i}\nu}(\mu \omega)
 \right)+\frac{\alpha}{\mu^2}L_s(1,\texttt{i}\nu,\mu \omega)\end{gather}
 where $\omega=\sqrt{x_0^2-x_3^2},\ \texttt{J}_{\mathbf{i}\nu}(\mu \omega)$ and
 $\texttt{Y}_{\mathbf{i}\nu}(\mu \omega)$ are Bessel functions of
 the first and second kind, $L_s(1,\texttt{i}\nu,\mu \omega)$ is the Lommel function s, $J_{\texttt{i}\nu}
 (\mu \omega)$ and $E_{\texttt{i}\nu}(\mu \omega)
 $ are Anger and Weber functions.

If $\mu\nu=0$ and $x_0^2>x_3^2$ then solutions of  (\ref{sol2}) are
reduced to the following form: \begin{gather}\label{niki}
\theta=c_7\sin(\nu\ln\omega) +c_8
\cos(\nu\ln\omega)+\frac{\delta}{\nu^2}
+\frac{\alpha\omega^2}{\nu^2+4}\ \ \texttt{if} \ \mu=0,
\nu\neq0;\\
 \label{som1}
\theta=\frac14\alpha\omega^2+\frac{\delta}2\ln^2(\omega)+c_7\ln(\omega)+c_8\
\ \texttt{if} \ \ \mu=\nu=0;\ \ \ \ \\
\label{som2}\theta=c_7\texttt{J}_0(\mu\omega)+c_8\texttt{Y}_0(\mu\omega)+
\frac{\alpha}{\mu^2} \ \ \texttt{if}\ \ \nu=\delta=0,\ \mu\neq0.
\end{gather}

We shall not present the cumbersome general solution of equation
(\ref{sol2}) for $x_0^2-x_3^2<0$ but restrict ourselves to the
particular case when $\alpha=\frac{\mu^2}{\nu^2}\delta$. Then
\begin{gather*}\label{uf}\theta=c_7\left(\texttt{I}_{\texttt{i}\nu}
(\mu\tilde\omega)
+\texttt{I}_{-\texttt{i}\nu}(\mu\tilde\omega)\right)+
c_8\left(\texttt{K}_{\texttt{i}\nu}(\mu\tilde\omega)
+\texttt{K}_{-\texttt{i}\nu}(\mu\tilde\omega)\right)+\frac{\delta}{\nu^2}
\end{gather*}
where  $\tilde\omega=\sqrt{x_3^2-x_0^2}.$

Algebra $ A_8: \quad \langle J_{12}, P_0, P_3 \rangle$,
\begin{gather*}
B_1=\frac{c_2 x_2 \theta+c_1 x_1-c_6 x_2}{\omega^2}, \
B_2=\frac{-c_2 x_1 \theta+c_1 x_2+c_6 x_1}{\omega^2}, \
B_3=-c_3 \theta +c_4,\\
E_1=\frac{c_1 x_1 \theta+c_5 x_1-c_2 x_2}{\omega^2}, \quad
E_2=\frac{c_1 x_2 \theta+c_5 x_2+c_2 x_1}{\omega^2}, \quad E_3=c_3
\end{gather*}
where $\omega^2=x_1^2+x_2^2$ and $\theta $ is a solution of equation
(\ref{sol2}) with \begin{gather}\label{be}\nu^2= {c_1^2-c_2^2},\ \
\mu^2={c_3^2-m^2},\ \ \delta=c_1c_5+c_2c_6,\ \
\alpha=c_3c_4.\end{gather}
 If  $
c_1^2\geq c_2^2$ and $c_3^2\geq m^2$  then $\theta$
 is defined
by relations (\ref{WEB})--(\ref{som2})  where $\mu,\nu$ and $\delta$
are constants given in (\ref{be}). If $c_1^2-c_2^2=-\lambda^2<0,\
m^2<c_3^2$ and $\alpha(\alpha\lambda^2+\delta\mu^2)=0$ then
\begin{gather*}\theta=c_7\texttt{J}_\lambda(\mu \omega)+c_8\texttt{Y}_\lambda(\mu
\omega)-\frac{\delta\pi}{2\lambda}\left(\cot
\left(\frac{\pi\lambda}{2}\right)J_\lambda(
 \mu \omega)+E_\lambda(\mu \omega)\right)+\frac{\alpha}{\mu^2}
\end{gather*} where $\texttt{J}_\lambda(\mu\omega)$ and $\texttt{Y}_\lambda(\mu\omega)$ are the Bessel
functions of the first and second kind, $J_{\lambda}
 (\mu \omega)$ and $E_{\lambda}(\mu \omega)$ are Anger and Weber functions
 correspondingly. In addition,
\begin{gather}\label{MU}
\theta=c_7\omega^{{\lambda}}+c_8\omega^{-{\lambda}}-\frac{\delta}{\lambda^2}
-\frac{\alpha}{\lambda^4},
\ \lambda^2={c_2^2-c_1^2}\ \text{ if }\ c_2^2>c_1^2, \ \ c_3^2=m^2;\\
\label{NU}\theta=c_7\texttt{I}_\lambda(\kappa\omega)+
c_8\texttt{K}_\lambda(\kappa\omega)+f \ \ \texttt{if}\ \
m^2-c_3^2=\kappa^2>0,\ c_2^2\geq c_1^2,
\end{gather}
where
\begin{gather}\label{bo}\begin{split}&f=-\frac{\delta}{\kappa^2}\
\texttt{ if }\ \delta=\alpha\frac{\kappa^2}{\lambda^2},\ \
\lambda\neq 0,\ \ f=\frac{4\alpha}{m^4x^2}-\frac{\alpha}{m^2}\ \
\texttt{if}\ \ \lambda=2,\ \delta=0,\\& f=-\frac{\alpha}{2\kappa}\
\texttt{ if }\ \delta=\lambda=0,\end{split}
\end{gather} $\texttt{I}_\lambda(\kappa\omega)$ and
$\texttt{K}_{\lambda}(\kappa\omega)$ are the modified Bessel
functions of the first and second kind.

Solutions (\ref{NU}) are valid also for parameters $\delta$ and
$\lambda$ which do not satisfy conditions presented in (\ref{bo}).
The corresponding function $f$ in (\ref{NU}) can be expressed via
the Bessel and hypergeometric functions, but we will not present
these cumbersome expressions here.

Algebra $ A_{19}: \quad \langle J_{12}, J_{03}, P_0-P_3 \rangle$
\begin{gather*},
B_1=E_2=\frac{c_1( x_1+ x_2 \theta)}{(x_3+x_0)(x_1^2+x_2^2)},\quad
B_2=-E_1=\frac{c_1( x_2-x_1 \theta)}{(x_3+x_0)(x_1^2+x_2^2)},\\
B_3=-c_3 \theta+c_2,\ \ E_3=c_3
\end{gather*}
where $\theta$ is a function of $\omega=\sqrt{x_1^2+x_2^2}$ which
solves equation (\ref{sol2}) with $\nu=\delta=0,\ \
\mu^2={c_3^2-m^2},\ \ \alpha=c_2c_3.$ Its explicit form is given by
equations (\ref{niki}) and (\ref{NU}) were $\delta=0$.

Algebra $ A_{24}: \quad \langle G_1, J_{03}, P_2 \rangle$,
\begin{gather*}
B_1=-x_3 \varphi,\quad B_2=-\frac{c_2x_0}{\omega^3}
-\frac{c_1}{x_0+x_3},\quad
B_3=x_1 \varphi,\\
E_1=-\frac{c_2x_3}{\omega^3} +\frac{c_1}{x_0+x_3},\quad E_2=x_0
\varphi,\quad E_3=\frac{c_2x_1}{\omega^3}
\end{gather*}
where $\omega=\sqrt{x_0^2-x_1^2-x_3^2}$, $\varphi=\varphi(\omega)$.
Functions $\varphi$ and $\theta$ should satisfy the following
equations:
\begin{gather*}
\omega\dot\varphi+3\varphi+\left(\frac{c_1}{\omega}+
\frac{c_2}{\omega^2}\right)\dot\theta=0,\ \
\ddot\theta+\frac{2\dot\theta}{\omega}-\left(c_1+\frac{c_2}
{\omega}\right)\varphi+m^2\theta=0.
\end{gather*}
If $c_1c_2=0$ then this system can be integrated in elementary or
special functions:
\begin{gather*}c_1=0:\ \ \ \varphi=-\frac{c_2\theta+c_3}{\omega^3};\\
\theta= c_4\sinh\frac{c_2}{\omega}+
c_5\cosh\frac{c_2}{\omega}\ \ \texttt{if}\
\ m=0,\ \ c_2\neq0,\\
\theta = \frac1\omega (c_4\sin m\omega+c_5\cos m\omega) \ \
\texttt{if}\ \ m\neq0,\ \ c_2=0,\ \
\texttt{and}\ \\
\theta=\frac{\texttt{D}}{{\omega}}\left(c_4+\int\frac{1}
{\texttt{D}^2\omega} \left(c_5+c_2c_3\int
\frac{\texttt{D}dx}{\omega^{5/2}}\right)dx\right)\ \ \texttt{if}\
m\neq0,\ \ c_2\neq0
\end{gather*}
where $\texttt{D}=\texttt{D}(0,m_-,n,m_+,f(\omega))$
 is the Heun double confluent
function with
\begin{gather*}m_\pm=m^2+c_2^2\pm\frac14,\ n=2(m^2-c_2^2),\
f(\omega)=\frac{\omega^2+1}{\omega^2-1}.
\end{gather*}
Let $c_2=0,\ c_1\neq0$, then
\begin{gather*} \varphi=\frac{1}{c_1}\left(\ddot
\theta+\frac{2\dot \theta}{\omega}+m^2\theta\right),\\
\theta=c_3G_1(c_1,\omega)+c_4(G_2(c_1,\omega)+G^*_2(c_1,{\omega}))+
\texttt{i}c_5(G_2(c_1,\omega)-G_2^*(c_1,{\omega}))\end{gather*}
where
\begin{gather*}G_1(c_1, \omega)=F\left(\frac{3+\texttt{i}c_1}2,
\frac{3-\texttt{i}c_1}2;\frac32;-\frac{m^2\omega^2}4\right),\\
G_2(c_1,\omega)=F\left(1+\texttt{i}c_1,1-\texttt{i}c_1;1+\frac{\texttt{i}c_1}2;
-\frac{m^2\omega^2}4\right)\omega^{-1+\texttt{i}c_1},\end{gather*}
$F(a,b;c;x)$ are hypergeometric functions and the asterisk denotes
the complex conjugation.

Algebra $ A_{27}: \quad \langle J_{03}+\alpha J_{12}, G_1, G_2
\rangle$
\begin{gather*}
B_1=\frac{\varphi_1}{x_0+x_3}+
\frac{(x_0+x_3)^2-x_1^2+x_2^2}{2(x_0+x_3)\omega^4}\varphi_3-
\frac{x_1x_2}{(x_0+x_3)\omega^4}\varphi_4,\\
B_2=\frac{\varphi_2}{x_0+x_3}+\frac{(x_0+x_3)^2+x_1^2-x_2^2}{2(x_0+x_3)
\omega^4}
\varphi_4-\frac{x_1x_2}{(x_0+x_3)\omega^4}\varphi_3,\\
E_1=-B_2+\frac{\varphi_3(x_0+x_3)}{\omega^4},\ \
E_2=B_1-\frac{\varphi_3(x_0+x_3)}{\omega^4},
\\E_3=\frac{x_2 \varphi_3-x_1 \varphi_4}{\omega^4},\ \ \
B_3=-\frac{x_1 \varphi_3+x_2 \varphi_4}{\omega^4},\\
\theta=\frac1\omega
\left(c_1\texttt{J}_1(m\omega)+c_2\texttt{Y}_1(m\omega)\right)
\texttt{ if }  m \neq0, \
\omega^2=x_0^2-x_1^2-x_2^2-x_3^2>0,\\\theta=\frac1{\tilde\omega}
\left(c_1\texttt{I}_1(m\tilde\omega)+c_2\texttt{K}_1(m\tilde\omega)\right)
\texttt{ if } m \neq0, \ \ \tilde\omega^2=-\omega^2>0,
 \\
  \theta= c_1+\frac{c_2}{\omega^2}\
\texttt{ if }\ m=0,\\ \varphi_1=c_2\cos{(\alpha
\ln{(x_0+x_3)})}+c_3\sin{(\alpha
\ln{(x_0+x_3)})},\\
\varphi_2=c_2\sin{(\alpha \ln{(x_0+x_3)})}-c_3\cos{(\alpha
\ln{(x_0+x_3)})},\\
\varphi_3=c_4\sin{\left(\alpha
\ln{\frac{\omega^2}{x_0+x_3}}\right)}+
c_5\cos{\left(\alpha \ln{\frac{\omega^2}{x_0+x_3}}\right)},\\
\varphi_4=c_4\cos{\left(\alpha
\ln{\frac{\omega^2}{x_0+x_3}}\right)}- c_5\sin{\left(\alpha
\ln{\frac{\omega^2}{x_0+x_3}}\right)},\
(c_3^2+c_2^2)(c_5^2+c_4^2)=0.
\end{gather*}
\subsection{Reductions to nonlinear ODE}
Using subalgebras $A_6,\ A_9,\ A_{10},\ A_{13},\ A_{14},\ A_{17}$
and $A_{18}$ we can reduce (\ref{kuriksha:PDE3}),
(\ref{kuriksha:PDE4}) to systems of ordinary differential equations
which however are nonlinear.

 Algebra $A_6: \quad \langle J_{03}+\alpha P_2, P_0, P_3 \rangle,\ \ \alpha\neq0$
 \begin{gather*}\bea{l} \displaystyle B_1=\varphi_1
\cosh{\frac{x_2}{\alpha}}-\varphi_2 \sinh {\frac{x_2}{\alpha}},\ \
B_2=\alpha \dot{\varphi}_2 \cosh{\frac{x_2}{\alpha}}-\alpha
\dot{\varphi}_1 \sinh {\frac{x_2}{\alpha}},\\ B_3=-c_1 \theta +c_2,
\\\displaystyle E_1=\alpha{\dot\varphi}_1\cosh{\frac{x_2}{\alpha}}-\alpha\dot\varphi_2\sinh
{\frac{x_2}{\alpha}}, \ \
E_2={\varphi}_1\sinh{\frac{x_2}{\alpha}}-\varphi_2\cosh
{\frac{x_2}{\alpha}},\ \ E_3=c_1\eea
\end{gather*}
where   $\theta,\ \varphi_1$ and $\varphi_2$ are
functions of $\omega=x_1$ which satisfy the following system of nonlinear
equations:
\begin{equation}\label{nl0}\alpha\dot\theta\varphi_2=
\alpha^2\ddot\varphi_2+\varphi_2,\
\varphi_1\dot\varphi_2-\dot\varphi_1\varphi_2=c_3,\end{equation} and
\beq\label{nl1}
\ddot\theta=(m^2-c_1^2)\theta+\alpha(\dot\varphi_1\varphi_1-
\dot\varphi_2\varphi_2)+c_1c_2.\eeq

We could find only particular solutions of this complicated system,
which correspond to some special values of arbitrary constants.
First let us present solutions linear in $\omega$:
\begin{gather}\label{fififi}\theta=\frac\omega\alpha-\frac{c_1c_2}{m^2-c_1^2}
\pm\mu c_5,\ \ \varphi_1=c_4\varphi_2,\ \
\varphi_2=\pm\frac{\omega}{\mu}+c_5, \  \ c_4^2\neq1, \ \ c_1^2\neq
m^2\end{gather} where $\mu=\sqrt{\frac{|m^2-c_1^2|}{|1-c_4^2|}}.$
 If
$c_1^2=m^2$ and $\varphi_1=\varphi_2$ then $\theta$ is given by
equation (\ref{A1c}) with $c=-c_1c_2$ while $\varphi_2$ is a linear
combination of Airy functions:
\begin{equation}\label{N02}\bea{l}\displaystyle\varphi_2 =
c_7\texttt{Ai}\left(\lambda(\omega-\nu)\right)
+c_8\texttt{Bi}\left(\lambda(\omega-\nu)\right)
 \eea\eeq where $\lambda=\left(\frac{c_1c_2}{\alpha}\right)^3,\
 \nu=\frac1{\alpha c_1c_2},\ \alpha c_1c_2\neq0$.

 If $c_1^2=m^2, c_2=0, c_3\neq0$ then we find a particular solution:
 \begin{gather*}\theta=\left(\alpha\mu^2+\frac1{\alpha}\right)x_1+c_5,\ \
  \varphi_2=c_6\cosh \mu x_1+c_7\sinh
 \mu x_1,\ \  \varphi_1=\frac1{\mu}\dot\varphi_2\end{gather*}
 where $\mu=\frac{c_3}{c_7^2-c_6^2}$ and $c_7^2\neq c_6^2$.

If $c_1^2=m^2$ and $\varphi_1=c_4\varphi_2$, $c_4\neq1, c_2=0$ then
\begin{gather}\label{113}\theta=\alpha\lambda\int{\varphi_2^2d\omega}+
\frac{c_5}{\alpha}\omega+c_6 \end{gather} where
$\lambda=\frac12(c_4^2-1)$ and $\varphi_2$ is an elliptic function
which solves the equation
\begin{gather}\label{114}\ddot
\varphi_2=\lambda\varphi_2^3-\kappa\varphi_2\end{gather} where
$\kappa=\frac{1-c_5}{\alpha^2}$. In addition, equation (\ref{114})
admits particular solutions in elementary functions:
\begin{gather}\label{115}\varphi_2=\pm\sqrt{\frac{\kappa}
{\lambda}} \tanh\left(\sqrt{\frac{\kappa}{2}}\omega+c_7\right)\ \
\texttt{if} \ \ c_5<1,\ \ c_4^2>1,\\\label{116}
\varphi_2=\pm\sqrt{\frac{\kappa} {\lambda}}
\tan\left(\sqrt{\frac{-\kappa}{2}}\omega+c_7\right)\ \ \texttt{if} \
\ c_5>1,\ \
c_4^2<1,\\\label{117}\varphi_2=\pm\frac{\sqrt{2}}{\sqrt{\lambda}\omega}\
\ \texttt{if}\ \ c_5=1, \ \ c_4>1.\end{gather}

If  $c_1^2=m^2+\frac1{\alpha^2}  \texttt{ and }\
\varphi_1=\pm\sqrt{1+c_4^2}\varphi_2$,  then we can set
\begin{gather*}\theta=\alpha c_4\varphi_2+c_1c_2\alpha^2\end{gather*}
and $\varphi_2$ should satisfy the following equation
\begin{gather*}\ddot \varphi_2-c_4\dot
\varphi_2\varphi_2+\frac{1}{\alpha^2}\varphi_2=0.\end{gather*} Its
solutions can be found in the implicit form:
\begin{gather*}\omega=c_4\alpha^2\int_0^{\varphi_2}\frac{dt}{W\left
(c_5^2\texttt{e}^{\frac12c_4^2\alpha^2t^2}\right)+1}+c_6\end{gather*}
where $W$ is the Lambert function, i.e., the analytical at $y=0$
solution of equation $W(y)\texttt{e}^{W(y)}=y$.

 Finally, for $c_1^2\neq m^2$ and $\varphi_1=\varphi_2$ we find
 the following solutions:
\beq\label{N03}\bea{l}\displaystyle\theta=\frac12\left(2\nu
c_5\sinh{2\nu \omega}+c_6\cosh{2\nu
\omega}\right)-\frac{c_1c_2}{4\nu^2}, \ \ 2\nu=\sqrt{m^2-c_1^2},
\\\displaystyle\varphi_2=\texttt{D}\left(0,m_+,n,m_-,\tanh \nu \omega\right)
\left(c_7+c_8\int\frac {dx_1}{\texttt{D}^2\left(0,m_+,n,m_-,\tanh
\nu \omega\right)}\right) \eea\eeq where
$\texttt{D}(0,m_+,n,m_-,\tanh \nu \omega)$ is the Heun double
confluent functions with
$m_\pm=\frac{c_5}{\alpha}\pm\frac1{\nu^2\alpha^2},\
n=\frac{c_6}{\alpha\nu}.$ If in (\ref{N03}) $c_5=\frac{c_6}{2\nu}$
and $\frac{c_6}{\nu \alpha}=-\frac{\kappa^2}{2}<0$
 then  \begin{gather}\label{N04}\varphi_2=c_7\texttt{J}_{\frac{\texttt{i}}{\nu\alpha}}
 \left(\kappa\texttt{e}^{\nu\omega}\right)+c_8\texttt{Y}_{\frac{\texttt{i}}{\nu\alpha}}
 \left(\kappa\texttt{e}^{\nu\omega}\right).\end{gather}

  Algebra $A_9: \quad \langle J_{23}+\alpha
P_0, P_2, P_3 \rangle,\ \ \ \alpha\neq0$
\begin{gather*}
E_1= c_1 \theta+c_2,\ \ E_2=\varphi_1
\cos{\frac{x_0}{\alpha}}-\varphi_2 \sin{\frac{x_0}{\alpha}},\ \
E_3=\varphi_1 \sin{\frac{x_0}{\alpha}}+\varphi_2
\cos{\frac{x_0}{\alpha}},\\ B_1=c_1, \ \ B_2=-\alpha \dot{\varphi}_1
\cos{\frac{x_0}{\alpha}}+\alpha \dot{\varphi}_2
\sin{\frac{x_0}{\alpha}}, \ \ B_3=-\alpha \dot{\varphi}_1
\sin{\frac{x_0}{\alpha}}-\alpha \dot{\varphi}_2
\cos{\frac{x_0}{\alpha}}
\end{gather*}
where $\varphi_1,\ \varphi_2$ and  $\theta$  are
functions of $\omega=x_1$ which satisfy equations (\ref{nl0}) and the following equation:
\begin{gather*}
\ddot\theta=(c_1^2+m^2)\theta-\alpha(\dot\varphi_1\varphi_1+
\dot\varphi_2\varphi_2)+c_1c_2.\end{gather*}

A particular solution of this system is $\varphi_1=c_4\varphi_2$ and
$\theta, \varphi_2$ given by equation (\ref{fififi}) where
$c_1^2\to-c_1^2$.

 If $c_1^2+m^2=0$ then we obtain solutions given
by equations (\ref{113}), (\ref{114}), (\ref{116}) (\ref{117}) where
$\lambda=-\frac12(c_4^2+1)$, and the following solutions:
\begin{gather}\label{j}\begin{split}&
\varphi_1=c_6\cos\mu\omega+c_7\sin\mu\omega,\ \
\varphi_2=c_6\sin\mu\omega-c_7\cos\mu\omega,\\&
\theta=\left(\frac1\alpha-
\alpha\mu^2\right)\omega+c_5\end{split}\end{gather} where
$\mu=\frac{c_3}{c_6^2+c_7^2}$.
\\\\ Algebra
$A_{10}: \quad \langle J_{12}+\alpha P_3, P_1, P_2 \rangle$,
$\alpha\neq0$
\begin{gather*}
B_1=\varphi_1 \cos{\frac{x_3}{\alpha}}-\varphi_2
\sin{\frac{x_3}{\alpha}},\quad B_2=\varphi_1
\sin{\frac{x_3}{\alpha}}+\varphi_2 \cos{\frac{x_3}{\alpha}} ,\quad
B_3=c_1,\\
E_1=\alpha \dot{\varphi}_1 \cos{\frac{x_3}{\alpha}}-\alpha
\dot{\varphi}_2 \sin{\frac{x_3}{\alpha}} ,\ \ E_2=\alpha
\dot{\varphi}_1  \sin{\frac{x_3}{\alpha}}+\alpha \dot{\varphi}_2
\cos{\frac{x_3}{\alpha}},\ \ E_3= c_1 \theta+c_2
\end{gather*}
where $\varphi_1, \varphi_2
$ and   $\theta$  are functions of $\omega=x_0$
satisfying (\ref{nl0}) and the following equation:
\begin{gather*}
\ddot\theta=-(c_1^2+m^2)\theta-\alpha(\dot\varphi_1\varphi_1+
\dot\varphi_2\varphi_2)-c_1c_2.\end{gather*} If $c_1^2+m^2=0$ we
again obtain solutions (\ref{j}) and solutions given by equations
(\ref{113}), (\ref{114}), (\ref{116}) (\ref{117}) where
$\lambda=-\frac12(c_4^2+1)$.

Algebra $ A_{17}: \quad \langle J_{03}+\alpha J_{12}, P_0, P_3
\rangle,\ \alpha\neq0$
\begin{gather*}
B_1=(\alpha \varphi_2'x_2 -\varphi_2x_1
)e^{-\frac{\zeta}{\alpha}-\omega}+(x_1 \varphi_1+\alpha
\varphi_1'x_2
)e^{\frac{\zeta}{\alpha}-\omega},\\
B_2=-(\alpha \varphi_2'x_1 +\varphi_2x_2)
e^{-\frac{\zeta}{\alpha}-\omega}-(\alpha \varphi_1'x_1
-\varphi_1x_2)e^{\frac{\zeta}{\alpha}-\omega},\quad
B_3=-c_1 \theta +c_2,\\
E_1=(\alpha \varphi_2'x_1 +\varphi_2x_2
)e^{-\frac{\zeta}{\alpha}-\omega}-(x_1 \alpha
\varphi_1'-\varphi_1x_2)
e^{\frac{\zeta}{\alpha}-{\omega}},\\
E_2=(\alpha \varphi_2'x_2 -\varphi_2x_1
)e^{-\frac{\zeta}{\alpha}-{\omega}}-(\alpha \varphi_1'x_2
+\varphi_1x_1)e^{\frac{\zeta}{\alpha}-{\omega}},\quad E_3=c_1,\
\end{gather*}
where $\omega=\frac12\ln(x_1^2+x_2^2)$, $\zeta=\arctan{\frac{x_2}{x_1}}$
. Functions $\varphi_1=\varphi_1(\omega),\ \varphi_2=\varphi_2(\omega)$ and $\theta=\theta(\omega)$
 should satisfy (\ref{nl0}) and the following
equation:
\begin{gather}\label{ufe}
\texttt{e}^{-2\omega}\ddot\theta+(m^2-c_1^2)\theta
+2\alpha(\dot\varphi_1\varphi_2+ \varphi_1\dot\varphi_2) +c_1c_2=0.
\end{gather}
This rather complicated system has the following particular
solutions for $  \ c_1=\pm m$:
\begin{gather}
  \theta=\frac1{a} {\omega}+c_4, \ \varphi_1={c_5},\
 \varphi_2={c_6};\label{x1}\end{gather}\begin{gather*}
 \theta=\left(\frac1{\alpha}+\alpha k^2\right)\omega+c_4,\ \
 \varphi_1=c_5\texttt{e}^{\kappa\omega},\
 \varphi_2=c_6\texttt{e}^{-\kappa\omega} \ \
 \texttt{if\ \ } 2c_5c_6k+c_3=0;\\
 \theta=-\frac14 c_1c_2\texttt{e}^{2\omega}+c_4\omega+c_5,\ \  \varphi_1=0,
 \ \varphi_2=c_6\texttt{J}_\mu\left(k\texttt{e}^\omega\right)+
 c_7\texttt{Y}_\mu\left(k\texttt{e}^\omega\right),\\
 \mu=\frac1\alpha \sqrt{c_4\alpha-1}\ \  \texttt{if}\
 \frac{c_1c_2}{2\alpha}=k^2>0,
 \end{gather*}
 and
 \begin{gather}\label{BIG}
\varphi_1=\kappa\varphi_2,\ \theta=\varphi_2\texttt{e}^{\omega}\ \
\varphi_2=2\mu\tan\left(\mu\texttt{e}^\omega\right)+c_4
\end{gather}
$\texttt{if}\ \kappa=\frac1{2\alpha},\  c_1c_2=4\mu^2>0,\
\alpha=\pm1.$ In (\ref{BIG}) we restrict ourselves to the particular
value of $\alpha$ in order to obtain the most compact expressions
for exact solutions.

An exact solution of equation (\ref{nl0}), (\ref{ufe}) for
$m^2-c_1^2=4\lambda^2>0$ and $c_2=0$ is given by the following
equation:
\begin{gather}\label{BIG1}\theta=\frac{\texttt{e}^{4\mu(1+\alpha^2)\omega+
2\lambda^2\texttt{e}^{2\omega}}}{\int{
\texttt{e}^{4\mu(1+\alpha^2)\omega+
2\lambda^2\texttt{e}^{2\omega}}d\omega}+c_4},\ \
\varphi_2=\theta\texttt{e}^\omega,\ \ \varphi_1=\mu\varphi_2
\end{gather}
where $\lambda,\ \mu$ and $\alpha$ are arbitrary real numbers.
\\

Algebra $ A_{18}: \quad \langle \alpha J_{03}+ J_{12}, P_1, P_2
\rangle,\ \alpha\neq0 $
\begin{gather*}
B_1=e^{-2\omega}\left(\left(\varphi_1x_0-\alpha\dot\varphi_2x_3\right)\cos\zeta-
\left(\varphi_2x_0+\alpha\dot\varphi_1{x_3}\right)\sin{\zeta}\right),\\
B_2=e^{-2\omega}\left(\left(x_0\varphi_1-\alpha\dot\varphi_2x_3\right)
\sin\zeta+
\left(x_0\varphi_2+\alpha\dot\varphi_1{x_3}\right)\cos{\zeta}\right),\ \ B_3=c_1, \\
E_1=e^{-2\omega}\left(\left(-\alpha\dot\varphi_2x_0+{\varphi_1}x_3\right)
\sin{\zeta}+\left(\alpha \dot\varphi_1x_0+
\varphi_2x_3\right)\cos{\zeta}\right),\\
E_2=e^{-2\omega}\left(\left(\alpha\dot\varphi_2x_0-{\varphi_1}x_3\right)
\cos{\zeta}+\left(\alpha \dot\varphi_1x_0+
\varphi_2x_3\right)\sin{\zeta}\right),\ \ E_3=c_1\theta-c_2
\end{gather*}
where $ \omega=\frac12\ln(x_0^2-x_3^2),\quad \alpha\zeta=
\ln{(x_0+x_3)}-\ln(x_0-x_3),$ $\varphi_1, \ \varphi_2$ and $\theta$
are functions of $\omega$ which should solve the system including (\ref{nl0}) and the following
equation:
\begin{gather*}
\texttt{e}^{-2\omega}\ddot\theta=-(m^2+c_1^2)\theta+{\alpha}(\dot\varphi_1\varphi_1+
\dot\varphi_2\varphi_2)+c_1c_2.\end{gather*} Particular solutions of
this system for $m^2+c_1^2=0$ are:
\begin{gather*} \theta=\left(\frac{1}{\alpha}-\alpha
k^2\right)\omega+c_4,\ \ \varphi_1=c_5\sin(k\omega),\
\varphi_2=c_6\cos(k\omega),\ \ \kappa=-\frac{c_3}{c_5c_6}.
\end{gather*}
In addition, the solutions (\ref{x1}), (\ref{BIG}) and (\ref{BIG1})
are valid where $\omega\to \ln(x_0^2-x_3^2).$
\subsection{Reductions to PDE}
Finally, let us make reductions of system (\ref{kuriksha:PDE3}),
(\ref{kuriksha:PDE4}) using the remaining subalgebras, i.e., $A_{6}$
with $\alpha=0$ and $A_{28}-A_{30}$. Basis elements of these
algebras do not satisfy condition (\ref{la}) and so it is not
possible to use the classical symmetry reduction approach. However,
to make the reductions we can impose additional conditions on
dependent variables which force equations (\ref{la}) to be
satisfied.

This idea is used in the weak
 transversality
 approach discussed in \cite{wint}. Moreover, in this approach
 the condition (\ref{la}) by itself is used  to find algebraic conditions
 for elements of matrices $\varphi^k_i$.

 We use even more week conditions which we call
 extra weak
 transversality. In other words we also look for additional
 constraints to solutions of equations (\ref{kuriksha:PDE3}),
 (\ref{kuriksha:PDE4}) which force condition (\ref{la}) to be
satisfied. But instead of the direct use of algebraic condition
(\ref{la}) we also take into account their differential
consequences. As a result we make all reductions for the system
(\ref{kuriksha:PDE3}), (\ref{kuriksha:PDE4}) which can be obtained
in frames of the weak
 transversality approach and also some additional reductions.

 Let us start with algebra $A_6$ with $\alpha=0$.
The set of the related basis elements $\langle P_0,\ P_3,\
J_{03}\rangle$ does not satisfies condition (\ref{la}). If we
consider this condition as an additional algebraic constraint for
solutions of equations (\ref{kuriksha:PDE3}), (\ref{kuriksha:PDE4})
then components of vectors $\mathbf{E}$ and $\texttt{B}$ should
satisfy the following conditions:
\begin{gather}\label{pde7}E_1=E_2=B_1=B_2=0\end{gather}

Substituting (\ref{pde7}) into (\ref{kuriksha:PDE3}),
(\ref{kuriksha:PDE4}) and supposing that $E_3, \ B_3$ and $\theta$
depend only on invariants $x_1$ and $x_2$  we can find the
corresponding exact solutions. However, we will obtain more general
solutions using the following observation.

To force condition (\ref{la}) to be satisfied it is possible to
apply additional conditions which are weaker than (\ref{pde7}). In
particular, we can ask for the following constrains:
\begin{gather}\label{wc}E_1=f_1(B_1,B_2),\quad
E_2=f_2(B_1,B_2)\end{gather} where $f_1(B_1,B_2)$ and $f_2(B_1,B_2)$
are differentiable functions of $E_1$ and $E_2$. These constrains
should be compatible with the field equations and make vanish the
term $E_1\partial_{B_2}-E_2\partial_{B_1}-
B_1\partial_{E_2}+B_2\partial_{E_1}$ in $J_{03}$. This term became
trivial provided
\begin{gather}\label{coco}f_1\frac{\p f_1}{\p B_2}-f_2\frac{\p f_1}{\p
B_1}+B_2=0;\quad f_1\frac{\p f_2}{\p B_2}-f_2\frac{\p f_2}{\p
B_1}-B_1=0.\end{gather}

The compatibility condition is much more complicated. However, it is
satisfied at least for linear functions $f_1(B_1,B_2)$ and
$f_2(B_1,B_2)$.

 Up to Lorentz transformations such constraints are
exhausted by the following ones:
\begin{gather}\label{6.1}E_1=0,\ E_2=0;\\ \label{6.2}E_1=0,\
B_2=0;\\ \label{6.3}E_1=B_1,\ E_2=B_2.\end{gather}

Let relations (\ref{6.1}) are fulfilled and $B_2, \ B_3,\ E_1,\
E_3,\ \theta$ depend on the invariant variables $x_1$ and $x_2$.
Then the system (\ref{kuriksha:PDE3}), (\ref{kuriksha:PDE4}) is
solved by the following vectors:
\begin{gather}\label{pd12}E_1=E_2=0,\ E_3=c_3, \
B_1=\frac{\partial \phi}{\partial x_2},\ B_2=-\frac{\partial
\phi}{\partial x_1},\ \quad B_3=-c_3\theta +c_2
\end{gather}
provided function $\theta=\theta(x_1,x_2)$ satisfies the following
equation
\begin{gather}\label{pd1}
\frac{\partial^2\theta}{\partial
x_1^2}+\frac{\partial^2\theta}{\partial
x_2^2}=(m^2-c_3^2)\theta+c_1c_2.
\end{gather}
 and $\phi=\phi(x_1,x_2)$ solves the two-dimension Laplace
equation:
\begin{gather}\label{pd11}\frac{\partial^2\phi}{\partial
x_1^2}+\frac{\partial^2\phi}{\partial x_2^2}=0.
\end{gather}

A particular solution of equation (\ref{pd1}) is:
\begin{gather}\label{OO}\begin{split}&
\theta=X(x_1)Y(x_2)+\frac{c_1c_2}{c_1^2-m^2}\ \texttt{ if }\
c_1^2\neq m^2,\\ &\theta=X(x_1)Y(x_2)+\frac{c_1c_2}{2}(x_1^2+x_2^2)
\ \texttt{ if }\ c_1^2= m^2\end{split}
\end{gather} where
\begin{equation}\label{OOO}\bea{l}
X(x_1)=c_{3,\mu}\mathbf{e}^{k_\mu x_1}+c_{4,\mu}\mathbf{e}^{-k_\mu
x_1},\ \ Y(x_2)=c_{5,\mu}\cos(n_\mu x_2)+c_{6,\mu}\sin(n_\mu x_2).
\eea\eeq
 Here  $k_\mu^2=m^2+\mu^2,\ n_\mu=c_1^2+\mu^2$, and $c_{s,\mu},\
s=3,4,5,6$ are arbitrary constants. The general solution of equation
(\ref{pd1}) can be expressed as a sum (integral) of functions
(\ref{OO}) over all possible values of $\mu$ and $c_{s,\mu}$.

Solutions (\ref{pd12}) include an arbitrary harmonic function
$\phi$. Only a very particular case of this solution corresponding
to $\phi=\texttt{Const}$ can be obtained in frames of the standard
weak
 transversality approach discussed in \cite{wint}.

Analogously, imposing condition (\ref{6.2}) we obtain the following
solutions:
\begin{gather}\label{pd3}E_1=0,\ E_3=c_1,\ B_2=0,\
B_3=-c_1\theta+c_2\end{gather} and
\begin{gather}\label{pd4}\begin{split}&
E_2=c_3\texttt{e}^{\frac1\mu(c_4\texttt{e}^{\mu x_2}
-c_5{\texttt{e}}^{-\mu
x_2})}+c_6\texttt{e}^{\frac1\mu(c_5\texttt{e}^{-\mu x_2}
-c_4{\texttt{e}}^{\mu
x_2})},\\&B_1=c_3\texttt{e}^{\frac1\mu(c_4\texttt{e}^{\mu x_2}
-c_5{\texttt{e}}^{-\mu
x_2})}-c_6\texttt{e}^{\frac1\mu(c_5\texttt{e}^{-\mu x_2}
-c_4{\texttt{e}}^{\mu x_2})},\\ &\theta=x_1(c_4\texttt{e}^{\mu x_2}
+c_5{\texttt{e}}^{-\mu x_2})+c_7\texttt{e}^{\mu x_2}
+c_8{\texttt{e}}^{-\mu x_2}+\frac{c_1c_2}{c_1^2-m^2}\\& \texttt{if
}\ c_1^2-m^2=\mu^2>0;
\\&E_2=c_3\texttt{e}^{\frac1\nu(c_4\cos{\nu x_2}
-c_5\sin{\mu x_2})}+c_6\texttt{e}^{-\frac1\nu(c_4\cos{\nu x_2}
-c_5\sin{\mu x_2})},\\& B_1=c_3\texttt{e}^{\frac1\nu(c_4\cos{\nu
x_2} -c_5\sin{\mu x_2})}-c_6\texttt{e}^{-\frac1\nu(c_4\cos{\nu x_2}
-c_5\sin{\mu x_2})},\\
&\theta=x_1(c_4\sin{\mu x_2} +c_5\cos{\mu x_2})+c_8\sin{\mu x_2}
+c_{9}\cos{\mu x_2}+\frac{c_6c_7}{c_6^2-m^2}\\& \texttt{ if }\
c_6^2-m^2=-\nu^2<0;
\\&E_2=c_3\sinh\left(\frac12c_4x_2^2+c_5x_2\right)+
c_6\cosh\left(\frac12c_4x_2^2+c_5x_2\right),\\
&B_1=c_6\sinh\left(\frac12c_4x_2^2+c_5x_2\right)+c_3\cosh
\left(\frac12c_4x_2^2+c_5x_2\right),\\&
\theta=x_1(c_4x_2+c_5)-\frac12c_1c_2x_2^2+c_7x_2+c_8\ \ \texttt{ if
}\ c_1^2=m^2.
\end{split}
\end{gather}
If conditions (\ref{6.3}) are imposed then one obtains the solutions
\begin{gather}\label{pd5}E_\alpha=B_\alpha=\partial_\alpha\phi,\
\alpha=1,2,\ E_3=c_1, \ B_3=-c_1\theta+c_2\end{gather} where $\phi$
is a function satisfying (\ref{pd11}), and $\theta$ is a solution of the following equation:
\begin{gather}\label{pd111}
\frac{\partial^2\theta}{\partial
x_1^2}+\frac{\partial^2\theta}{\partial
x_2^2}=(m^2-c_1^2)\theta+c_1c_2+(\p_1\phi)^2+(\p_2\phi)^2.
\end{gather}
Another solution corresponding to (\ref{6.3}) is given by equations
(\ref{pde7}) for $B_1, B_2, E_1, E_2$ and $E_3=c_1,\ B_3=-c_1\theta+c_2$, while
$\theta$ is given by equation (\ref{OO}).

 Algebra $A_{28}: \ \
 \langle G_{1},G_{2},J_{12}\rangle$
\begin{gather*}
B_1=E_2=\frac{1}{(x_0+x_3)^3}\left(c_1x_1-x_2(c_1\theta+c_2)\right),\\
B_2=-E_1=\frac{1}{(x_0+x_3)^3}\left(c_1x_2+x_1(c_1\theta+c_2)\right),
\
\\B_3=\frac{c_1}{(x_0+x_3)^2},  \quad E_3=-\frac{(c_1\theta
+c_2)}{(x_0+x_3)^2},\ \ \theta=\frac{\varphi}{x_0+x_1}\ \
\end{gather*}
where $\varphi$ is a function of two variables
$\omega=\frac{x_0^2-x_1^2-x_2^2-x_3^2}{2(x_0+x_3)}$ and
$\zeta=x_0+x_3$, which  satisfies the following equation:
\begin{gather}\label{Noo}
\frac{\partial^2\varphi}{\partial\omega\partial\zeta}=
\left(\frac{c_1^2}{\zeta^4}-m^2\right)\varphi+\frac{c_1c_2}{\zeta^3}.
\end{gather}
Let $m=c_1=0$ then $\varphi=\varphi_1(\omega)+\varphi_2(\zeta)$
where $\varphi_1$ and $\varphi_2$ are arbitrary functions. For
$c_1=0, m^2\neq0$ equation (\ref{Noo}) admits solutions in separated
variables:
\begin{gather}
\varphi=\sum_\mu( a_\mu\sin(\nu_\mu\xi_+)\sin(\mu\xi_-)+
b_\mu\cos(\nu_\mu\xi_+)\cos(\mu\xi_-)\nonumber\\
+c_\mu\cos(\nu_\mu\xi_+)\sin(\mu\xi_-)+
d_\mu\sin(\nu_\mu\xi_+)\cos(\mu\xi_-))\label{NNN}
\end{gather}where
 $\xi_\pm=\omega\pm\zeta,\ \
\ \ \nu_\mu^2=m^2+\mu^2$ and $\mu, S_\mu, a_\mu, b_\mu, c_\mu$ and
$d_\mu$ are arbitrary constants.

For $c_1\neq0$ we obtain:
\begin{gather*}
\varphi=\frac{c_1c_2\zeta}{c_1^2-m^2\zeta^4}+ \sum_\mu
R_\mu\texttt{e}^{\mu\omega-\frac{3m\zeta^4+c_1^2}{3\mu\zeta^3}}.\end{gather*}
Algebra $ A_{29}:\ \ \langle J_{01},J_{02},J_{12}\rangle $
\begin{gather*}
B_1= \frac{x_2(c_1 \theta+c_2)}{\omega^3}, \quad B_2=- \frac{x_1(c_1
\theta+c_2)}{\omega^3},
\quad B_3= \frac{c_1x_0}{\omega^3},\\
E_1=-\frac{c_1x_2}{\omega^3}, \quad E_2=\frac{c_1x_1}{\omega^3},
\quad E_3=\frac{x_0(c_1 \theta+c_2)}{\omega^3}, \quad
\theta=\frac{\varphi}{\omega}
\end{gather*}
where $\omega^2=x_0^2-x_1^2-x_2^2$ and $\varphi$ is a function of
$\omega$ and $x_3$ which satisfy the following equation:
\begin{gather}\label{007}
\frac{\partial^2\varphi}{\partial
x_3^2}-\frac{\partial^2\varphi}{\partial \omega^2}=\left(\frac{c_1^2
}{\omega^4}+m^2\right)\varphi+\frac{c_1c_2}{\omega^3}\ \texttt{ if }
\ x_0^2>x_1^2+x_2^2
\end{gather}
where $\omega=\sqrt{x_0^2-x_1^2-x_2^2}$, and
\begin{gather}\label{0071}
\frac{\partial^2\varphi}{\partial
x_3^2}+\frac{\partial^2\varphi}{\partial
\tilde\omega^2}=\left(m^2-\frac{c_1^2
}{\tilde\omega^4}\right)\varphi-\frac{c_1c_2}{\tilde\omega^3}\
\texttt{ if } \ x_0^2<x_1^2+x_2^2
\end{gather}
where $\tilde\omega=\sqrt{x_1^2+x_2^2-x_0^2}$.

 Let $c_1=m=0$ then the general solution of equation (\ref{007}) is:
 $\varphi=\varphi_1(\omega+x_3)+\varphi_2(\omega-x_3)$
 where $\varphi_1$ and $\varphi_2$ are arbitrary functions. Solutions which
 correspond to
 $c_1=0, m\neq0$ can be obtained from (\ref{NNN}) by changing
 $\xi_+\to x_3,\ \xi_-\to \omega$.  If $c_1\neq 0$ and $m\neq 0 $ then
 \begin{gather}\nonumber
\varphi=\sum_\mu \texttt{D}_\mu\left(\left(a_\mu+b_\mu\int
 \frac{d\omega}{\omega \texttt{D}_\mu^2}\right)\sin(\mu x_3)
  + \left(c_\mu+d_\mu\int
 \frac{d\omega}{\omega
 \texttt{D}_\mu^2}\right)\cos(\mu x_3)\right)\\\label{002}
 -c_1c_2\int\left(\frac1{\omega
 \texttt{D}_0^2}\int\frac{\texttt{D}_0d\omega}{\omega^{5/2}}\right)d\omega
\end{gather}
where $\texttt{D}_\mu=\texttt{D}(0,k^-_\mu,s,k^+_\mu,f(\omega))$ is
the double confluent Heun function with
$k^\pm_\mu=m^2-\mu^2+c_1^2\pm\frac14, \ s=2(m^2-\mu^2-c_1^2),\
f(\omega)=\frac{\omega^2+1}{\omega^2-1}, \ \mu, a_\mu, b_\mu, c_\mu$
and $d_\mu$ are arbitrary constants.

Solutions of equation (\ref{0071}) also can be represented in the
form (\ref{002}) where $\omega\to\tilde\omega$ and
\begin{gather*}k^\pm_\mu=\mu^2-m^2+c_1^2\pm\frac14, \ s=2(\mu^2-m^2-c_1^2),\
f(\omega)=\frac{\tilde\omega^2+1}{\tilde\omega^2-1}.\end{gather*}

Algebra $ A_{30}:\ \langle J_{12},J_{23},J_{31}\rangle $
\begin{gather*}
{B}_a=\frac{c_1{x}_a}{r^3}, \quad {E}_a=\frac{(c_1 \theta-c_2)
{x}_a}{r^3},\ \ \theta=\frac{\varphi}{r},
\end{gather*}
where $r=\sqrt{x_1^2+x_2^2+x_3^2}$ and $\varphi$ is a function of
$r$ and $x_0$ satisfying the following equation:
\begin{gather}\label{LL}
\frac{\partial^2\varphi}{\partial
r^2}-\frac{\partial^2\varphi}{\partial
x_0^2}=\left(\frac{c_1^2}{r^4}+m^2\right)\varphi-\frac{c_1c_2}{r^3}.
\end{gather}

Solutions of this equation can be represented in the form
(\ref{002}) where $\omega=r$, $x_3\to x_0$ and $\texttt{D}_\mu=\texttt{D}(0,k^-_\mu,s,k^+_\mu,f(\omega))$ is
the double confluent Heun function with
$k^\pm_\mu=-(m^2+\mu^2+c_1^2)\pm\frac14, \ s=2(c_1^2-m^2-\mu^2),\
f(\omega)=f(r)=\frac{r^2+1}{r^2-1}$.

A special solution of equation (\ref{LL}) corresponding to $c_2=0$
and zero constant of variable separation is given in (\ref{ha}).

\subsection{Solutions with maximal number of arbitrary elements}

Solutions considered in the above include arbitrary parameters and
in some cases even arbitrary functions. At the end of our analysis a
special class of solutions will be presented which depend on six (!)
arbitrary functions. This class cover all reductions which can be
obtained using subalgebras $A_{12}-A_{14}$ and $A_{20}-A_{23}$.

Let us define
\begin{gather}\label{ar1}\begin{split}&B_1=E_2=\psi_1(x_1,x_2,\omega)
-x_1\dot\varphi_1(\omega)-x_2\left(\dot\varphi_4(\omega)-
\varphi_5(\omega)\dot\theta\right),\\&
B_2=-E_1=\psi_2(x_1,x_2,\omega)-x_2\dot\varphi_5(\omega)+x_1\left(\dot\varphi_2(\omega)-
\varphi_1(\omega)\dot\theta(\omega)\right),\\&
B_3=\varphi_1(\omega)+\varphi_5(\omega),\quad
E_3=\varphi_2(\omega)+\varphi_4(\omega)\end{split}\end{gather} where
$\varphi_1$,... $\varphi_5$ and $\psi_1, \psi_2$ are functions of
 $\omega=x_0+x_3$ and $x_1, x_2, \omega$ respectively, and
\begin{gather}\theta=-
\frac{(\varphi_1+\varphi_5)(\varphi_2+\varphi_4)}{m^2}, \texttt{ if
} F=-m^2\theta,\ m^2\neq0,\nonumber \\
\theta=\varphi_3(\omega),\quad
(\varphi_1+\varphi_5)(\varphi_2+\varphi_4)=0 \ \texttt{if }
F=0.\label{ar2}\end{gather}

Up to restriction present in (\ref{ar2}) functions $\varphi_1,
\varphi_2$ and $\varphi_3$ are arbitrary while $\psi_1$ and $\psi_2$
should satisfy the Caushy--Rieman condition with respect to
variables $x_1$ and $x_2$:
\begin{gather}\p_1\psi_1+\p_2\psi_2=0,\ \ \p_1\psi_2-\p_2\psi_1=0.
\label{ar3}\end{gather}

It is easy to verify that functions (\ref{ar1}) and (\ref{ar2}) do
solve equations (\ref{kuriksha:PDE3}), (\ref{kuriksha:PDE4}).

\section{Discussion}

The main goal of the present paper was  to find families of exact
solutions of field equations of axion electrodynamics using their
symmetry w.r.t. the Poincar\'e group P(1,3). To achieve this goal we
classify and find all possible reductions of these equations which
can be made using the three parameter subgroups of P(1,3). The
complete list of reductions which can be made using invariants of
these subgroups together with the obtained solutions are presented
in sections 3 and 4. Among them there are solutions including sets
of arbitrary parameters and arbitrary functions as well. In
addition, it is possible to generate more extended families of exact
solutions applying the inhomogeneous Lorentz transformations to the
found ones.

For such subalgebras whose basis elements do not satisfy the
transversality condition (\ref{la}) we apply the week and "extra
weak" transversality approach, see section 4.4. As a result we find
solutions (\ref{pd12})--(\ref{pd3}) which cannot be found applying the
standard weak transversality conditions discussed in \cite{wint}.

Making reductions of equations (\ref{kuriksha:PDE}),
(\ref{kuriksha:PDE5}) we restrict ourselves to functions $F$ linear
in $\theta$. However, these reductions do not depend of the choice
of $F$; to obtain reduced equations with $F$ arbitrary it is
sufficient  simple to change $m^2\theta\to -F(\theta)$ or even
$m^2\theta\to -F(\theta,p_\mu p^\mu)$ everywhere.

Except a particular example given by relations
(\ref{N20})--(\ref{N115}) we did not present exact solutions for
equations (\ref{eq3}). Let us note that reductions of these
equations can be made in a very straightforward way. Indeed, making
the gauge transformation $E_a\to\texttt{e}^{\theta}E_a$ and
$B_a\to\texttt{e}^{\theta}B_a$ we can reduce these equations to a
system including the Maxwell equation for the electromagnetic field
in vacua and the following equation:
\begin{gather}\label{endfin}\Box \theta =\kappa (\textbf{B}^2 -
\textbf{E}^2)\texttt{e}^{-2\theta}+F.\end{gather} Since reductions
of the free Maxwell equations with using three-dimension subalgebras
of p(1,3) have been done in paper \cite{lahno}, to find the related
exact solutions for system (\ref{eq3}) it is sufficient to solve
equation (\ref{endfin}) with $\mathbf{B}$ and $\mathbf{E}$ being
exact solutions found in \cite{lahno}.

Solutions presented in sections 3 and 4 can have various useful
applications. Indeed, the significance of exact solutions, even
particular ones, can be rather high. First they present a certain
information about particular properties of the model. Secondly, they
can solve an important particular boundary value problem, a famous
example of this kind is the Barenblat solution for the diffusion
equation \cite{Barenblat}. In addition, the particular exact
solutions  can be used to test the accuracy of various approximate
approaches.

The found solutions, especially those which include arbitrary
functions or, like  (\ref{wond1}),  satisfy the superposition
principle, are good candidates to applications in various initial
and boundary value problems of axion electrodynamics. Some of these
solutions, e.g., (\ref{bed}), (\ref{A1a}) with $\mu=1$ and
$c_1^2+c_2^2>1$, describe the wave propagation with the group
velocity higher then the velocity of light. Moreover, these
solutions are smooth and bounded functions which correspond to
positive definite and bounded energy density \cite{NK1}. In spite of
that these solutions are causal since their energy velocity does not
exceed the velocity of light \cite{NK1}.

The existence of linear solutions for nonlinear partial differential
equations is a very interesting phenomenon which helps to clarify
basic properties of some nonlinear models. We have indicated some
particular solutions of axion electrodynamics which satisfy the
superposition principle. A systematical study of such solutions for
the Hirota bilinear equations was carried out in paper \cite{Ma1}.

We believe that the list of exact solutions presented in sections 3
and 4 can find other interesting applications. In particular,
solutions, which correspond to algebras $A_9$, $A_{1}$, $A_{17}$,
$A_{18}$ and $A_{28}$ generate well visible dynamical contributions
to the axion mass. In addition, as it was indicated in \cite{ninni},
the vectors of the electric and magnetic fields described by
relations (\ref{N11}) give rise to exactly solvable Dirac equation
for a charged particle anomalously interacting with these fields. We
plane to present a detailed analysis of the obtained solutions
elsewhere.

We find a number of exact solutions
which can be obtained using Lie symmetries of the axion
electrodynamics, and only few more general solutions, see (\ref{ar6}) and (\ref{ar1})--(\ref{ar2}). Of course there are also other approaches to symmetry
analysis and construction of exact solutions of partial differential
equations. Additional tools for construction of exact
solutions are presented by  Lie-B\"acklund symmetries first used by Emma Noether as long as in 1918.

Let us mention some of more modern approaches: nonclassical method of
Bluman and Cole \cite{Blum} which is also known as the "method of
conditional symmetries" \cite{FuFu}, \cite{wwint},  the direct method of Clarkson
and Kruskal \cite{Clark}, potential symmetries (see, e.g., \cite{pop}).
We remind that the powerful inverse problem method is based on the infinite number of conservation
 laws \cite{mar}, etc. And it is the whole series of possible symmetries which  exhibits  integrability of partial differential equations, see \cite{Ma2} for discussion of this point.

 Of course it would be interesting to apply the generalized symmetries to construct exact solutions of the axion electrodynamics. Since the Lie-B\"aclund symmetries and the related conservation laws for the Maxwell equations in vacua have been described in paper \cite{mesh}, there is a good starting point for their application to the non-linear system  (\ref{kuriksha:PDE}), (\ref{kuriksha:PDE5}) and especially to system (\ref{eq3}). However, in the present paper we restrict ourselves to completed description of the solutions which can be obtained using the three-dimensional subalgebras of p(1,3), keeping in mind that one day it would be possible to obtain more general classes of solutions.

\begin{center}\large Acknoledgments\end{center}
We would like to thank Professors Roman Popovych and Stephen Anco
for useful discussions.

\end{document}